\documentclass[conference,10pt]{IEEEtran}  

\usepackage{array}
\usepackage{amssymb}
\usepackage{amsmath}
\usepackage{graphicx}
\usepackage{xcolor}
\usepackage{subfig}
\usepackage{boxedminipage}
\usepackage{hyperref}
\usepackage[ruled,vlined]{algorithm2e}
\usepackage{algorithmic}
\usepackage{balance}
\usepackage{amsmath}
\usepackage{xy}
\xyoption{all}

\newcommand{\eat}[1] {}

\newcommand{\calC} {{\cal C}}

\title{Efficient Inferencing of Compressed Deep Neural Networks}

\author{
	\IEEEauthorblockN{
		Dharma Teja Vooturi \IEEEauthorrefmark{1},
		Saurabh Goyal\IEEEauthorrefmark{2}
		Anamitra R. Choudhury\IEEEauthorrefmark{2},
		Yogish Sabharwal\IEEEauthorrefmark{2},
		Ashish Verma\IEEEauthorrefmark{2}
	}\vspace{1mm}
	\IEEEauthorblockA{
		\hspace{1mm}
		\IEEEauthorrefmark{1}
		IIIT Hyderabad, India \hspace{9mm}
		Email: dharmateja.vooturi@research.iiit.ac.in
	}
	\IEEEauthorblockA{
		\hspace{30mm}
		\IEEEauthorrefmark{2}		
		IBM Research - India\hspace{10mm}
		Email: \{sgoyal30, anamchou, ysabharwal, vashish\}@in.ibm.com
	}
}

\begin{document}

\maketitle              

\begin{abstract}
Large number of weights in deep neural networks makes the models difficult to be deployed in low memory environments such as, mobile phones, IOT edge devices as well as ``inferencing as a service" environments on cloud. Prior work has considered  reduction in the size of the models, through compression techniques like pruning, quantization, Huffman encoding etc. However, efficient inferencing using the compressed models has received little attention, specially with the Huffman encoding in place.  In this paper, we  propose efficient parallel algorithms for inferencing of single image and batches, under various memory constraints.  
Our experimental results show that our approach of using variable batch size for inferencing 
achieves 15-25\%   performance improvement  in the inference throughput for AlexNet, 
while maintaining  memory and latency constraints.

\end{abstract}

\section{Introduction}
Deep neural networks have been used extensively over the last decade in applications ranging from computer vision~\cite{KrizhevskySH12} to speech recognition~\cite{GravesS05} and natural language processing~\cite{Collobert2011}.
In this paper, we focus particularly on convolutional neural networks (CNNs) which have 
become ubiquitous  in object recognition, image classification, and retrieval.  Almost all of the recent successful recognition systems  are built on top of this architecture (see \cite{JiaSDKLGGD14, DonahueJVHZTD14, GongWGL14, ZeilerF14}).
A simple convolution neural network consists of a sequence of layers, with every layer of the network transforming one volume of activations to another through a differentiable function. 
Thus for an image classification problem, the  input image is transformed layer by layer from the original pixel values to the final class scores. 

Convolution networks comprise of different types of layers including convolution (CONV), fully connected layer (FC), pooling layer (POOL),  Rectified Linear Unit (ReLU)  etc. 
Of these, the CONV and the FC layers contain weights and biases, which are   parameters trainable over some data set.
Thus the  CONV/FC layers perform transformations that are  functions of not only the activations in the input volume, but also of the parameters of the respective layers (the weights and biases of the neurons). On the other hand, the RELU/POOL layers  implement  fixed functions.

As datasets increase in size, so do the  number of layers in the CNNs and the number of parameters, in order to absorb the enormous amount of supervision. In 1998 Lecun et al. designed a CNN model LeNet-5 with less than 1M parameters to classify handwritten digits~\cite{Lecun98}, while in 2012, Krizhevsky et al.~\cite{KrizhevskySH12} won the ImageNet competition with 60M parameters and 8 layers (this correspond to the popular AlexNet network). Deepface classified human faces with 120M parameters~\cite{TaigmanYRW14}, and Coates et al.~\cite{CoatesHWWCN13} scaled up a network to 10B parameters. Karen Simonyan and Andrew Zisserman ~\cite{SimonyanZ14a} developed VGG-16 network with 16 layers
and 138M parameters.

The large number of weights in powerful and complex neural networks makes the models difficult to be deployed 
in low memory environments such as, mobile phones, IOT edge devices etc.
Large networks do not fit in on-chip storage  and are stored in external DRAM: thus they 
need to be  fetched every time for the inferencing of each test sample. This leads to multiple issues. Firstly, the inference time shoots up due to the overhead of external memory accesses. Secondly,
fetching the model from the external DRAM consumes large amount of energy. For instance,  Han et al.~\cite{HanMD15}
states that the energy cost per  fetch ranges from 5pJ for 32bit coefficients in on-chip SRAM to 640pJ for 32bit coefficients in off-chip
LPDDR2 DRAM:  thus running a 1 billion connection neural network,  at 20Hz would require  12.8W just for DRAM access - 
this prohibits inferencing on a typical mobile device.
A similar issue arises for ``inferencing as a service'' environment on the cloud: 
in this case the networks need to be loaded before the inferencing, thus increasing 
memory requirement and cost. It is therefore advisable to have the model reside permanently in the memory. Moreover many applications like visual recognition 
require multiple models for inferencing: thus it is feasible to have all the models loaded
apriori in memory only if the models are pretty small in size.

To address the above issues, significant work has been done to reduce the size of the networks.  
The objective in the ideal scenario is to generate a model of smaller size, with limited loss of accuracy in the prediction, and no sacrifice in the inference time.
Model compression can be  effected  through a combination of pruning, weight sharing and encoding of the connection weights.
In the pruning step,  the network is pruned  by removing the redundant connections of the network.
Next, the weights are quantized so that multiple connections share the same weight, thus only the codebook (effective weights) and the indices need to be stored. 
The codebook is generally of small size, and hence the indices can be represented with fewer bits than that required for the original weights.
Finally, some encoding (like Huffman coding) is performed to take advantage of the biased distribution of effective weights in further reducing the model size.

Neural network pruning has been pioneered even in the early development of neural networks (see ~\cite{Reed93}), and has been implemented through various strategies over the years.
Anwar et  al.~\cite{AnwarHS17} and Molchanov et  al.  ~\cite{MolchanovTKAK16} employ structured pruning  at the level of feature maps and kernels. The advantage of this 
scheme of pruning is that the resultant connection matrix can be considered dense. However their strategy is more suited for 
convolution layers. Song Han et  al.~\cite{HanMD15} have considered the weight based pruning where they remove all connections whose weights are lower than fixed threshold. 
Their pruning strategy (along with quantization and Huffman encoding) was able to get the model size of AlexNet reduced  from 240MB to 6.9MB, 
and that of VGG-16 from 552MB to 11.3MB, without loss of accuracy on Imagenet dataset.
A lot of literature is available for weight sharing and quantisation as well. 
Half-precision networks (Amodei et al., ~\cite{AmodeiABCCCCCCD15}) cut sizes of neural networks in half. XNOR-Net (Rastegari et al., ~\cite{RastegariORF16}), 
DoReFa-Net (Zhou et al., ~\cite{ZhouNZWWZ16}) and network binarization (Courbariaux et al.~\cite{CourbariauxB16}; Lin et al.~\cite{LinCMB15}) use aggressively quantized weights, activations and gradients to further reduce computation during training, however, the extreme compression rate comes with a loss of accuracy. Hubara et al.~\cite{HubaraCSEB16} and Li  Liu~\cite{LiL16} propose ternary weight networks to trade off between model size and accuracy.
Zhu et  al.~\cite{ZhuHMD16}  propose trained ternary quantization which uses two full-precision scaling coefficients for each layer, where these coefficients are trainable parameters.
Gong. et  al.~\cite{GongLYB14} consider vector quantization methods for compressing the parameters of CNNs.
HashedNets~\cite{ChenWTWC15} uses hash function to randomly group connection weights into hash buckets, so that all connections within the same hash bucket share a single parameter value.

In this work,  we consider the compression strategy as suggested by  Han et  al. (see ~\cite{HanMD15, HanPTD15}). As stated before, their compression technique has gained significant popularity due to the very little loss in accuracy for a number of networks and datasets. Since 
all connections with weights below a threshold are removed from the network, the pruned network becomes a 
sparse structure that is stored using compressed sparse row (CSR) or compressed sparse column (CSC) format.
The model is further compressed by storing the  index difference instead of the absolute position, and encoding this difference in 
$k$ bits for each layer: for an index difference larger than $2^k$,  zero padding is employed. 
Finally Huffman encoding is employed on both the weight clusters and the index differences to ensure that more common cluster centres and  index differences
are represented with fewer bits.

The real challenge with compressed models lies in processing them for inferencing.
Efficient inferencing using the compressed models has received little attention.
As stated before, with pruning the matrix becomes sparse and the indices are stored in the form of relative differences. 
With weight sharing,  a short (8-bit) index for each weight is stored. More indirection is added with Huffman encoding.
All of these increase the complexity of the inferencing process, making it inefficient on CPUs and GPUs.
The trivial method of uncompressing the matrix back to dense form and performing the inferencing in a standard framework
(like Caffe, Tensorflow etc) is not a good choice because of the 
excessive memory usage and the running time. 
Previous work has considered hardware and software accelerators to facilitate computation on compressed models.
Han et  al.~\cite{HanLMPPHD16} has proposed EIE, an efficient inference engine, that performs
customized sparse matrix vector multiplication and handles weight sharing with no loss of efficiency. 
However this requires  specialized hardware to be designed  to act as the accelerator. 
On the software side,
Intel Math Kernel Library (MKL~\cite{mkl}) provides  optimized sparse solvers for matrix-matrix and matrix-vector multiplications.
However it does not incorporate relative indexing and Huffman encoding, which are necessary to gain the desired compression levels.
Tensorflow  has recently incorporated Gemmlowp library (see ~\cite{gemmlowp}) for fast inferencing using 
eight-bit arithmetic rather than floating-point - however, this does not handle pruned and encoded models.

Another important aspect is the transition of  mobile devices  to multi-core CPUs. 
The current generation of mobile processors are being designed to deal with the increased number of high performance use cases. 
To satisfy the rapidly growing demand for performance and form factor sleekness, the industry has begun to adopt Symmetrical
Multiprocessing even on mobile systems. This calls for leveraging
multiple cores to facilitate faster inferencing even in low memory systems. 
Nvidia has studied and developed GPU-Based Inferencing  (see \cite{2015GPUBasedDL});
in a recent work Huynh et al.~\cite{Huynh2017} has proposed 
DeepMon, a mobile GPU based deep learning inference system for mobile devices. However all of these
work on uncompressed models. Thus
very little has been studied on parallel domain for compressed (in particular encoded) models.  


Another key factor is the batch size that should be used for inferencing on these limited resource systems.
It is well-known that larger batch size for inferencing increase both the throughput (since computing resources can be utilized more efficiently) and the latency. Thus inferencing applications strive to 
maximize a usable batch size while keeping latency under a given threshold.
The  maximum batch size is also determined by the amount of the available memory in the system. However, this varies dynamically depending on the system load.
Hence the  batch size for achieving the maximum throughput can only be figured out at the time of inferencing.

The focus of this paper is to study and propose optimizations for efficient inferencing compressed models under various resource limitations. 
Our main contributions are as follows:
\begin{itemize}
\item We propose a framework for inferencing of images with compressed models that rely on pruning, quantization, relative indexing and encoding techniques for compression. To the best of our knowledge, this is the first effort towards efficient inferencing using compressed models under memory constraints.
\item We propose parallel algorithms under this framework that can use tuned math libraries available on the platform to perform  efficient inferencing. Our framework uses different blocking schemes to optimize the inferencing time, wherein the best choice of the block size
depends on the layer of the network, its sparsity and the batch size used. 
\item We show that variable size batching that performs inferencing on
a different number of activations in each layer can lead to better inferencing performance. To this effect, we develop a novel dynamic programming based algorithm to figure out the optimal batch size to be used in the inferencing for each individual layer under memory and latency constraints.
\item Our experimental results show that our approach of using variable batch size for inferencing 
achieves 15-25\%   performance improvement  in the inference throughput for AlexNet, while maintaining  memory and latency constraints.
\end{itemize}


The rest of the paper is organized as follows.
In Section~\ref{motivation}, we motivate our problem by defining the challenges and the use cases. 
Section~\ref{sec:prelims} establishes  necessary preliminaries and concepts before we present our 
inferencing schemes in Section~\ref{sec:inference}. Our results for different blocking schemes are presented in Section~\ref{sec:expt1}.  We next study 
variable size batching for  inferencing in Section~\ref{sec:dp}, the results of which are
presented in Section~\ref{sec:expt2}. Finally, we conclude in Section~\ref{sec:conc}.

\section{Discussion on use cases and challenges}
\label{motivation}
Today, a large number of Artificial Intelligence (AI) applications rely on using deep learning models for various tasks, such as, image classification, speech recognition, natural language understanding, natural language generation and so on.
Due to the significant improvement in performance achieved by the deep learning models, there is a natural trend to use these models on the applications running on mobile phone and other edge devices in the context of IOT (Internet of Things). 
For example, more and more people now want to take pictures using their mobile phones and get information on the building and surroundings around them in a foreign place. Usage of voice based assistants on mobile phones and other home devices is another increasing trend. 
Applications in the area of augmented reality involves continuous image recognition with results being reported on a VR display to provide more information regarding the environment to the individual. 
For example, in security, this can be used for identity detection.
Similarly, in self-driven cars, deep learning models are used
to inference in real-time using data collected from a combination of sensing technologies including visual sensors, such as cameras, and range-to-object detecting sensors, such as lasers and radar. 
Increased instrumentation in various industries such as agriculture, manufacturing, renewable energy and retail generates lot structured and unstructured data which preferably needs to be analyzed at the edge device and so that real-time action can be taken.

For the scenarios described above, inferencing can be done either on the cloud (or server) or on the edge device itself. However, offloading  inferencing to  the cloud can be impractical in lot of situations due to  wireless  energy  overheads, turn-around latencies and data security reasons. On the other hand, given the sheer size of the deep learning models, inferencing on mobile/edge devices poses other kind of challenges on resources, such as memory, compute and energy which need to be utilized efficiently while continuing to provide high accuracy and similar latency.


Even when inferencing is done on the cloud, resources have to be efficiently utilized to keep the cost of inferencing minimum for the cloud vendor as the cost of inferencing is directly dependent on resource utilization. Just as an example, a vendor providing "Inferencing as a service" for image classification may want to keep hundreds of deep learning models customized for various domains and users in memory in order to provide the low response time. This calls for storing compressed models in-memory and directly inferencing using the compressed model when the requests come in. All of this has to be done without compromising on the latency and accuracy of the inferencing.


\section{Preliminaries}
\label{sec:prelims}
\subsection{Inferencing as matrix computations}

A fully-connected (FC) layer of a deep neural network (DNN) performs the computation as
\begin{equation}
b = Wa+v.
\label{eqn:fc1}
\end{equation}
where $a$ and $b$ are respectively the input activation vector and the output activation vector, $v$ is the bias, $W$ is the weight matrix.
The output activations of Equation ~\ref{eqn:fc1} is computed element-wise as:
\begin{equation}
b_i = \sum_{j=0}^{n-1}W_{ij}a_j+v_i.
\label{eqn:fc2}
\end{equation}

For a typical FC layer like FC7 of VGG-16 or AlexNet, the activation vectors are 4K long, and the weight matrix is 4K x 4K (16M weights). 
Weights are represented as single-precision floating-point numbers so such a layer requires 64MB of storage. Similarly for FC6 layer of 
AlexNet the weight matrix is of dimension 4096 x 9216, for  FC6 layer of 
VGG-16 the weight matrix is of dimension 4096 x 25088.

The computation of a convolution (CONV) layer of a CNN can also be expressed as a matrix-matrix multiplication operation.
The input activation for the CONV layer is a 3-dimensional tensor.
The convolution layer's parameters consist of a set of learnable filters (or kernels), which have a local connectivity 
along width and height  in the input,
but extend through the full depth of the input volume. Each filter is convolved across the width and height of the input volume, computing the dot product between the entries of the filter with the input and producing a 2-dimensional activation map of that filter. 
Stacking the activation maps for all filters along the depth dimension forms the full output volume of the convolution layer. 
The dot products between the filters and local regions of the input, can be formulated as a matrix-matrix multiplication, by 
flattening out the local regions of the input to individual columns
and the layer weights to rows: the result of a convolution is now equivalent to performing one large matrix which evaluates the dot product between every filter and every receptive field location.
See~\cite{deep_learn}
for details.


\subsection{Representation of compressed model}
As stated before,  this paper considers the compression technique, as suggested by  Han et  al.~\cite{HanMD15, HanPTD15}
to reduce the  size of the DNNs without loss of accuracy, obtained through a combination of pruning, weight sharing and Huffman encoding.
Pruning makes weight matrix $W$ sparse:
the pruned matrix $W$ is stored  in a variant of the standard compressed row storage (CSR) format.
The standard  CSR representation works as follows:
instead of storing the entire matrix $W$ of dimension $r$ x $c$,  
vectors, one of  floating-point numbers ({\em val}), and the other two of integers ({$col\_ind$, $row\_ptr$) are used. The $val$ vector stores the values of the nonzero elements of $W$, as they are traversed in a row-wise fashion.  The $row\_ptr$ vector stores the locations in the $val$ vector that start a row, that is, if $val(m)=W_{ij}$ then $row\_ptr(i) \leq m < row\_ptr(i+1)$. By convention, we define $row\_ptr(r+1) =nnz+1$, where $nnz$ is the number of nonzeros in $W$. The $col\_ind$ vector is used to store the column indexes of the elements in the val vector.   
 Figure~\ref{fig:repr_f2} shows the CSR representation for the matrix given in 
Figure~\ref{fig:repr_f1}.

\begin{figure*}[!tbp]
  \centering
  \subfloat[Sparse Matrix.]{\includegraphics[width=1.7in]{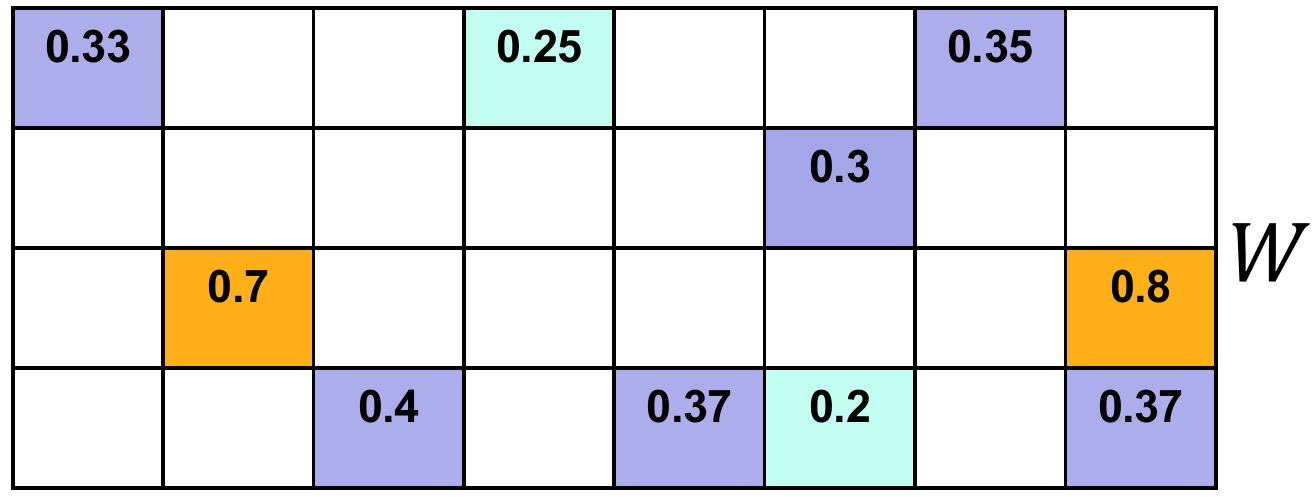}\label{fig:repr_f1}}
  \hfill
  \subfloat[CSR Format.]{\includegraphics[width=1.7in]{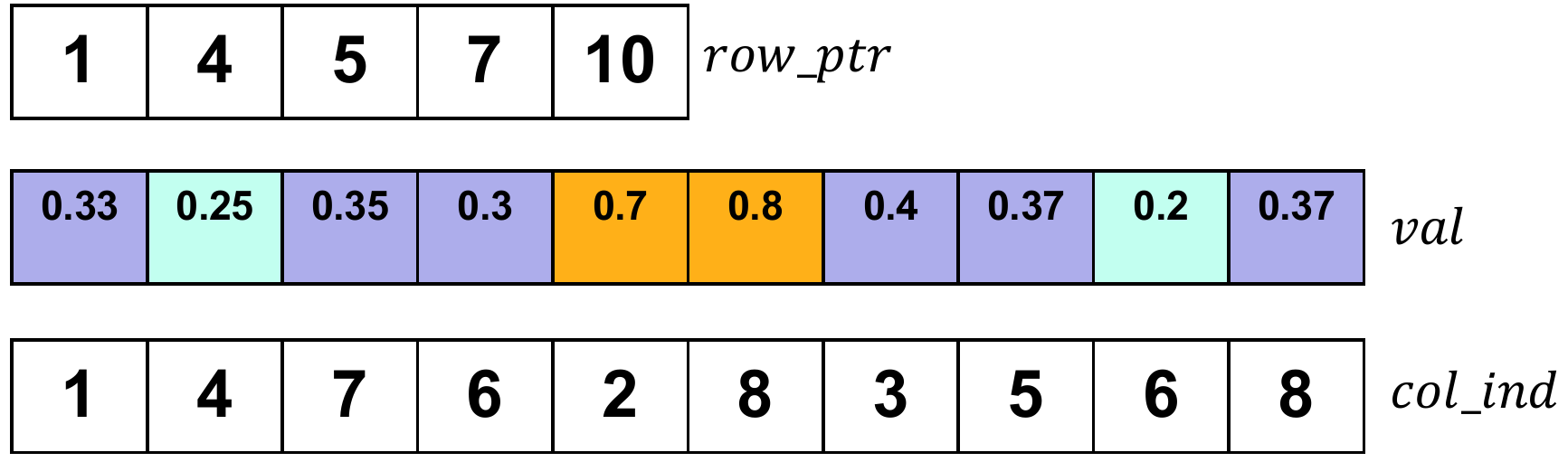}\label{fig:repr_f2}}
  \hfill
  \subfloat[Relative Column Index.]{\includegraphics[width=1.7in]{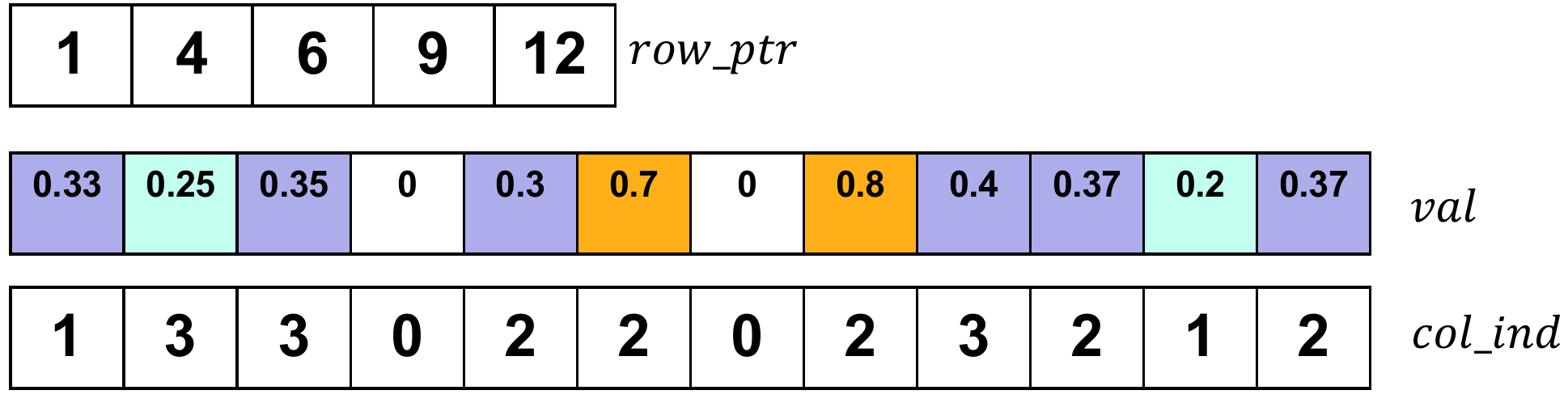}\label{fig:repr_f3}}
  \hfill
  \subfloat[Quantized Weight.]{\includegraphics[width=1.7in]{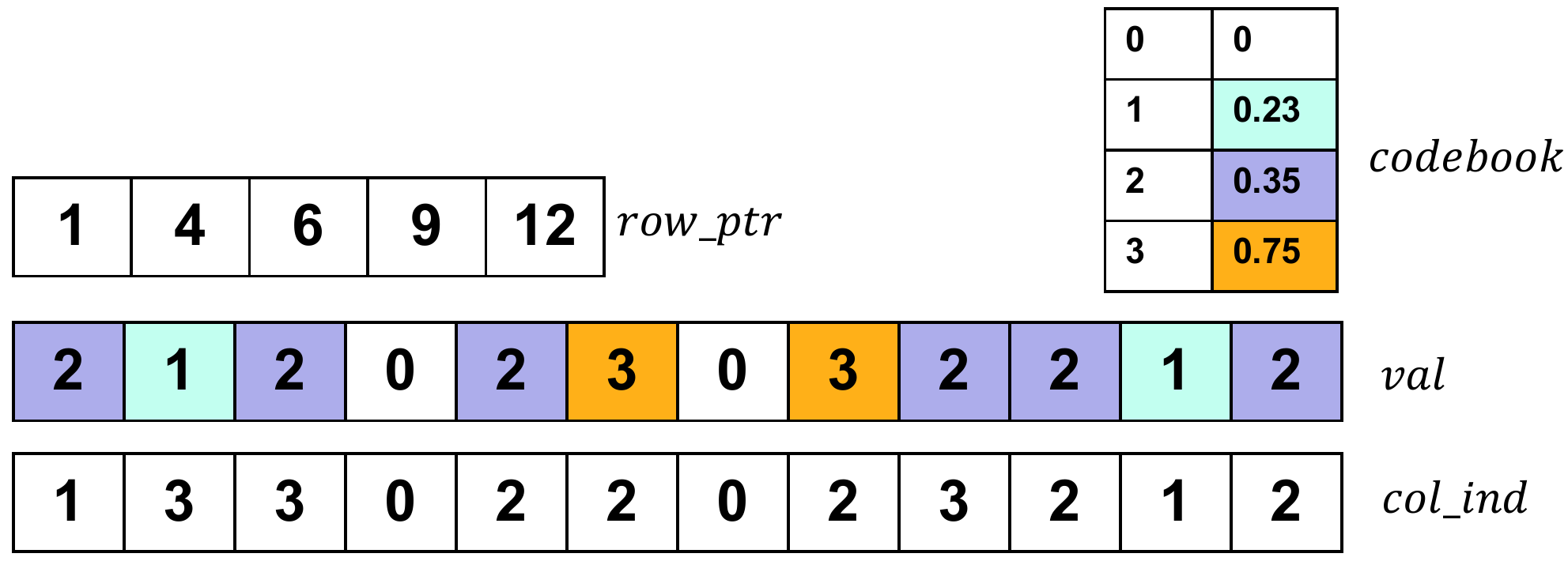}\label{fig:repr_f4}}
  \hfill
  \subfloat[Huffman Encoded Model.]{\includegraphics[width=2.8in]{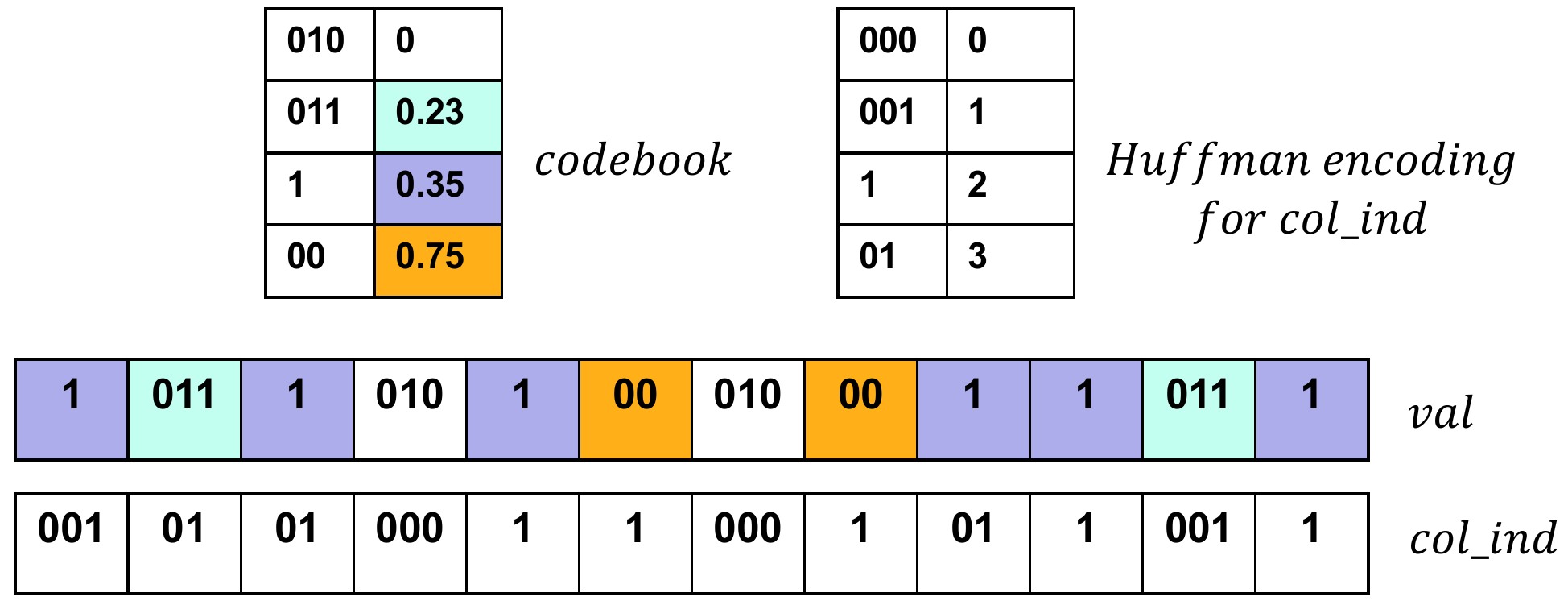}\label{fig:repr_f5}}
  \caption{Representation of a compressed model.}
\end{figure*}

The $col\_ind$ vector can be further compressed by making each of its entries exactly $k$ bits. This is achieved by modifying   $col\_ind$ as follows:
if $val(i)$ is the first non-zero entry of any particular row, then $col\_ind(i)$ is set to the corresponding column number; else 
$col\_ind(i)$ is set to the  number of columns between the current non-zero and the last non-zero entry of the row.
If more than $2^k$ zeros appear before a non-zero entry, we add a zero in both the $val$ and the $col\_ind$ vectors.
This representation format is the CSR with relative column index.
Figure~\ref{fig:repr_f3} shows the relative indexed CSR of  Figure~\ref{fig:repr_f2} with $k=2$. Since the first non-zero column of second row exceeds 4, 
we pad a zero at the fourth location
of both $val$ and $col\_ind$.  
Further compression is effected using quantization where similar valued non-zero entries of $val$ are clustered together to share the same 
value. If $r$ bits are used for quantization, we use at most ($2^r-1$) distinct  non-zero values along with 0 in quantized values, and  each entry of the $val$ vector is  a $r$ bit index
to the corresponding cluster centre. The cluster centre values are stored in the codebook. 
Figure~\ref{fig:repr_f4} shows the quantized model representation (quantization here is done using $2$ bits),  entries 
denoted by the same colour in the matrix of  Figure~\ref{fig:repr_f1} are mapped to a single cluster,  and each entry of the $val$ vector is  a $2$ bit index
to the corresponding cluster centre. The  codebook is also shown. 
Finally Figure~\ref{fig:repr_f5} shows the Huffman encoded bit representations of $val$ and $col\_ind$ vectors. 
Clearly,  entry $i$ of the $ row\_ptr$ array will a 2-tuple, the first field storing the starting address for row $i$ in $val$, while the 
second stores  the starting address for row $i$ in 
$col\_ind$.


\section{Inferencing using Compressed Models}
\label{sec:inference}

In this section, we discuss the various approaches for inferencing using the compressed model,
where the compressed model is stored  in  the format as shown in the previous section.
Clearly, the trivial method of exploding the model back to the dense format  and doing the computation 
(using standard frameworks like Caffe, Tensorflow etc) is not a good choice since the entire purpose of
model compression gets defeated because  of the 
excessive memory usage. The other extreme of decoding element by element of the matrix and doing the operations on the
decoded element 
has little memory overhead, but is computationally inefficient.
This calls for the need to develop an efficient stand-alone module (independent of the
Caffe/Tensorflow framework) for inferencing using the compressed model.
The na{\"i}ve algorithm for doing the inferencing is presented in Algorithm~\ref{alg:pseudocode1}.
The idea here is to work  sequentially on the individual rows of the weight matrix (line 3).
For a particular row, the $col\_ind$ and the $val$ entries for that row are first  Huffman-decoded (line 5-6);
this is followed by converting relative column index of $col\_ind$ to absolute index (line 7) 
and creating an $abs\_val$ array which  is essentially the $val$ array with its entries replaced by the
corresponding codebook entires. 
All these steps  in fact  create the arrays in Figure~\ref{fig:repr_f1} from 
Figure~\ref{fig:repr_f5} for a particular row segment.
Finally we call MKL routine $mkl\_scsrmm$ for 
sparse matrix-matrix multiplication  of $abs\_val(i)$ and $a$
to compute $b[i,:]$.


\begin{algorithm}[t]
\caption{Na{\"i}ve algorithm for inferencing using compressed model }
\label{alg:pseudocode1}
\small
\begin{algorithmic}[1]
	\STATE Input: $row\_ptr$ array, entry $i$ of which is a tuple 
	of  starting address of row $i$ in $val$ and that in $col\_ind$.\\
	$val$ Huffman encoded cluster index bit stream. \\
	$col\_ind$ Huffman encoded rel. indexed column bit stream. \\
	$\calC$ codebook of quantized weights. \\
	$a$ input activation matrix. \\
	\STATE Output: $b$ output  activation  matrix. \\
		 
	\FOR{every entry $i$ of the $row\_ptr$ array}
		\STATE Set $val\_begin(i)$,  $val\_end(i)$, $col\_begin(i)$, $col\_end(i)$\\
		 for row $i$ as follows \\
		 \quad  \quad $\langle val\_begin(i), col\_begin(i) \rangle \leftarrow row\_ptr(i)$\\
		 \quad  \quad $\langle val\_end(i), col\_end(i) \rangle \leftarrow row\_ptr(i+1)$.
		 \STATE $dec\_val(i)$ $\leftarrow$ Huffman decoding of bit stream in $val$ between $val\_begin(i)$ and  $val\_end(i)$.
		\STATE  $dec\_col(i)$ $\leftarrow$ Huffman decoding of bit stream in \\
		$col\_ind$ between $col\_begin(i)$ and  $col\_end(i)$.
		\STATE $abs\_col(i)$ $\leftarrow$  Prefix sum of $dec\_col(i)$.
		\STATE Set $abs\_val(i)[j]$ $\leftarrow$ $\calC [dec\_val(i)[j]]$ , $\forall j$.
		\STATE $b[i, :]$ += MKL\_CSRMM($abs\_val(i)$, $a$)
	\ENDFOR	
\end{algorithmic}
\end{algorithm}

The above algorithm can be parallelized by employing different threads to operate on different rows of the weight matrix.
Moreover MKL internally can use multiple threads for sparse matrix operations.
However  Algorithm~\ref{alg:pseudocode1} faces multiple drawbacks.
Firstly, the algorithm decodes an entire row of the matrix, and thus the memory requirement becomes
significant for large matrices. 
Secondly, most  algorithms for matrix multiplication work more efficiently using  blocks rather than individual elements, to achieve necessary reuse of data in local memory.
The advantage of this approach is that the small blocks can be moved into the fast local memory and their elements can then be repeatedly used.
This motivates us to employ blocking even for compressed model inferencing, which we describe next.

\subsection{Blocking of Weight Matrix}

 The
general idea of blocking is to organize the data structures in a program into  chunks called blocks. The program is
structured so that it loads a block into the L1 cache, does all the reads and writes that it needs to on that
block, then discards the block, loads in the next block, and so on. 
Similar to standard matrix multiplication, the blocking algorithm for inferencing shall work 
 by partitioning the matrices into submatrices and then exploiting
the mathematical fact that these submatrices can be manipulated just like scalars.
Instead of storing the  original weight matrix in row major format,
we need to ensure that any particular block of the matrix is stored in contiguous memory.  
This will make certain	 that the Huffman decoding happens on contiguous memory locations and  generates the submatrix 
corresponding to a block.

See Figure~\ref{fig:block_f1} and Figure~\ref{fig:block_f2} for illustration.
Suppose the original weight matrix  stored in dense row major format is of dimension 8x8, and we decide to work on blocks each sized 4x4. 
We first convert  this matrix to   4 x 16 format, such that each row of the new matrix stores 
elements of the corresponding block of the old matrix in contiguous locations. This new matrix 
is then stored in CSR format with relative indexing and Huffman encoding, as discussed in the 
previous section. 
\\
{\it Size of the modified model}:

It is observed that the non zeroes  in the weight matrix are uniformly distributed, thus
the size of the $val$ and $col\_ind$ vectors does not change a lot  (even with zero padding in the compressed format)  when the matrix is
stored in block contiguous fashion.
The number of rows in the modified matrix is same as the number of blocks in the original matrix, and may be larger or smaller than that
in the  original matrix depending on the block size. From experimental results, it is however observed, that
change in model size due to this difference in the size of the $row\_ptr$ is insignificant. Hence we can assume that 
storing the model in block contiguous fashion does not add to memory overhead.

\begin{figure}[!tbp]
  \centering
  \subfloat[Original Connection Matrix.]{\includegraphics[width=2in]{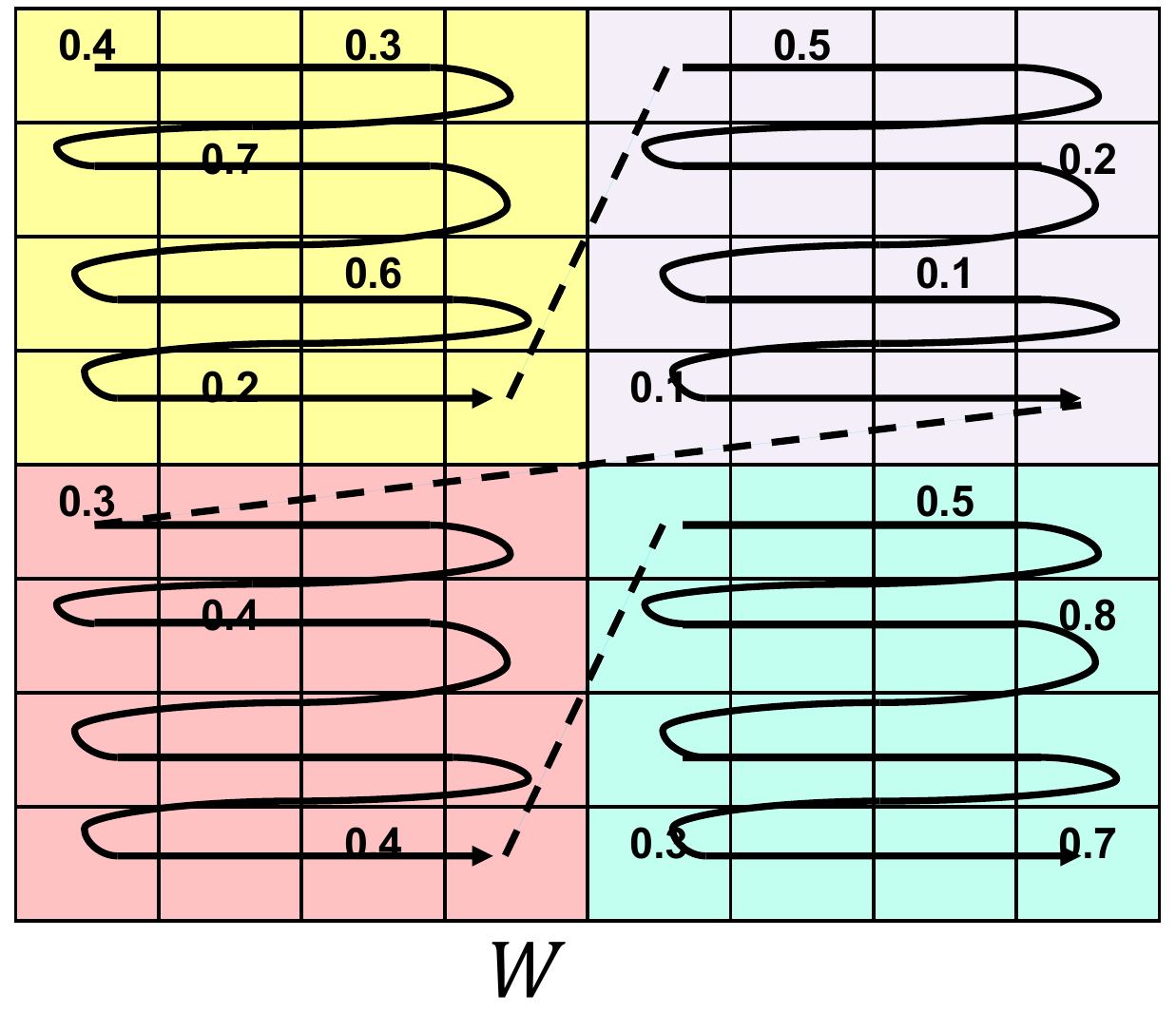}\label{fig:block_f1}}
 \hspace{10mm}
  \subfloat[Modified Connection Matrix]{\includegraphics[width=2.5in]{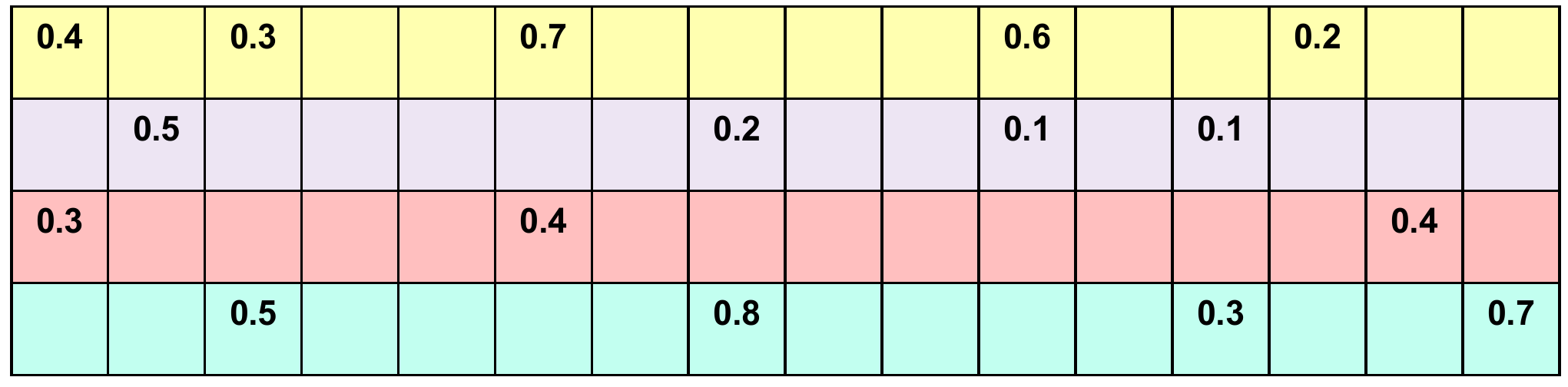}\label{fig:block_f2}}
  \caption{Representation of a compressed model.}
\end{figure}

\subsection{Blocked Inferencing Procedure}

Next we present our inferencing algorithm using the blocked storage scheme. 
Our algorithm ensures that once a  row of the connection matrix (which corresponds to a block
in the original weight matrix) is decoded,
the decoded entries are used for all the computations that require them.
This is illustrated in Figure~\ref{fig:block_mult}. A row is decoded and multiplied with all possible subblocks of 
input activation matrix to generate partial results for the output activation matrix. 
The blocked inferencing algorithm is presented in Algorithm~\ref{alg:pseudocode2}.

\begin{figure}[b]
\centering
\includegraphics[width=3in]{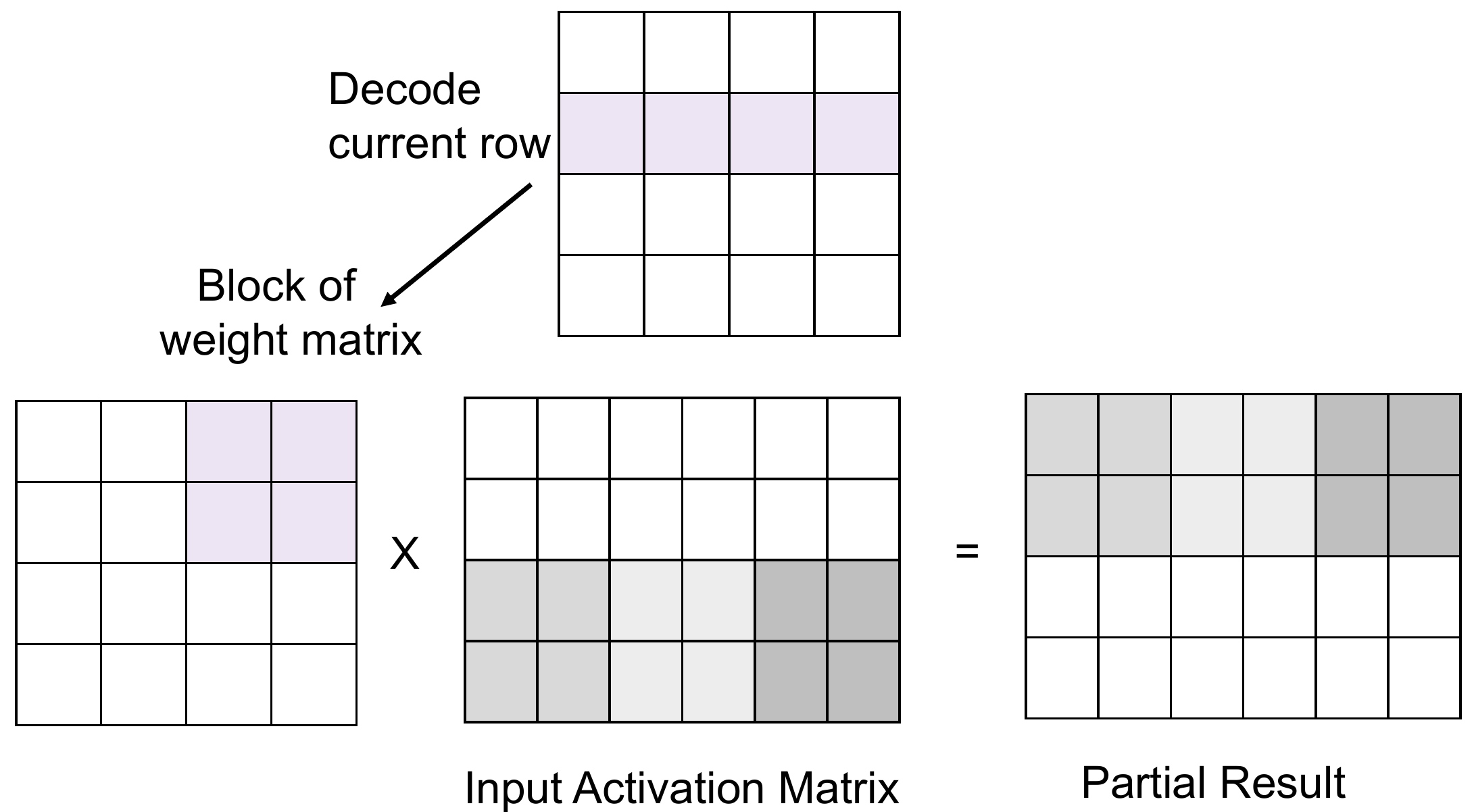}
\caption{Blocked inference scheme.}
\label{fig:block_mult}
\end{figure}

\begin{algorithm}[t]
\caption{Algorithm for block inferencing}
\label{alg:pseudocode2}
\small
\begin{algorithmic}[1]
	\STATE Input: Compressed model stored in $bh$ x $bw$ block contiguous manner with \\
	$row\_ptr$ array, entry $i$ of which is a tuple $\langle x,y \rangle$ \\
	where $x$ and $y$ being respectively starting address of row $i$ in $val$ and 	that in $col\_ind$.\\
	$val$ Huffman encoded cluster index bit stream. \\
	$col\_ind$ Huffman encoded relative indexed column bit stream. \\
	$\calC$ codebook of quantized weights. \\
	$a$ input activation matrix with $a_{rows}$ rows\\
	\STATE Output: $b$ output  activation matrix\\
		 
	\FOR{every entry $i$ of the $row\_ptr$ array}
		\STATE Set $val\_begin(i)$,  $val\_end(i)$, $col\_begin(i)$, $col\_end(i)$\\
		 for row $i$ as follows \\
		 \quad  \quad $\langle val\_begin(i), col\_begin(i) \rangle \leftarrow row\_ptr(i)$\\
		 \quad  \quad $\langle val\_end(i), col\_end(i) \rangle \leftarrow row\_ptr(i+1)$.
		\STATE $dec\_val(i)$ $\leftarrow$ Huffman decoding of bit stream in $val$ between $val\_begin(i)$ and  $val\_end(i)$.
		\STATE  $dec\_col(i)$ $\leftarrow$ Huffman decoding of bit stream in \\
		$col\_ind$ between $col\_begin(i)$ and  $col\_end(i)$.
		\STATE $abs\_col(i)$ $\leftarrow$  Prefix sum of $dec\_col(i)$.
		\STATE Set $abs\_val(i)[j]$ $\leftarrow$ $\calC [dec\_val(i)[j]]$ , $\forall j$.
		\STATE Arrange $abs\_val(i)$ as $bh$ x $bw$ block. \\
		\STATE col\_id = $ ( i \% (a_{rows}/bw) ) * bw$  \\
		\STATE row\_id = $ ( i / (a_{rows}/bw) ) * bh$  \\
		\STATE b[row\_id:(row\_id+bh-1),:] += MKL\_CSRMM($abs\_val(i)$, a[col\_id:(col\_id+bw-1),:] ) \\
	\ENDFOR	
\end{algorithmic}
\end{algorithm}

\section{Experimental Results with Blocking}
\label{sec:expt1}
In this section, we present the experimental results for our block inferencing procedure. We begin by specifying the system configurations and the dataset.

\subsection{System and Dataset}

For running our experiments (also the ones in Section~\ref{sec:expt2}), we have used Intel Xeon CPU E5-2697 system. It has two NUMA nodes 
with 12 cores, each with frequency of 2.70GHZ. The system has 32KB, 256KB and 30MB of L1, L2 and L3 cache respectively.
We consider compressed models for two popular deep neural networks, AlexNet and VGG-16. 
For each of these models we consider the compressed configurations corresponding to four different pruning percentages.
The first configuration corresponds to the procedure applied in  \cite{HanMD15}.  
Table~\ref{tab:alex_pr} and Table~\ref{tab:vgg_pr}
 present the pruning percentages of all the layers in this configuration. We refer to this configuration as 
{\em conventional} in subsequent discussion.
The compressed model sizes of AlexNet and VGG-16 for this configuration are respectively 6.81 MB and 10.64 MB.
The other three configurations correspond respectively to 70\%, 80\% and 90\% pruning of {\em all} the layers of the network.  
We consider these configurations to study how our scheme performs for a wide range of sparsity spectrum of the compressed models.
8 bit (5 bit) quantization for CONV (FC) layers
and 4 bit (5 bit) relative indexing for AlexNet (VGG-16) is employed for all the configurations.

\begin{table}[!tbp]
  \centering
  \subfloat[AlexNet]{
\begin{tabular}{|c|c|}
			\hline 
			Layer & Pruning \% \\ \hline 
			conv1 & 16 \\ \hline 
			conv2 & 62 \\ \hline 
			conv3 & 65 \\ \hline 
			conv4 & 63 \\ \hline 
			conv5 & 37 \\ \hline 
			fc6 & 91  \\ \hline 
			fc7 & 91  \\ \hline 
			fc8 & 75  \\ \hline
        \end{tabular}
        \label{tab:alex_pr}
  }
  \hspace{1mm}
    \subfloat[VGG-16]{
\begin{tabular}{|c|c|m{2em}|m{2em}|}
			\hline 
			Layer & Pruning \% \\ \hline 
			conv1\_1 & 42  \\ \hline
			conv1\_2 & 78 \\ \hline
			conv2\_1 & 66 \\ \hline
			conv2\_2 & 64 \\ \hline
			conv3\_1 & 47 \\ \hline
			conv3\_2 & 76 \\ \hline
			conv3\_3 & 58 \\ \hline
			conv4\_1 & 68 \\ \hline
			conv4\_2 & 73 \\ \hline
			conv4\_3 & 66 \\ \hline
			conv5\_1 & 65 \\ \hline
			conv5\_2 & 71 \\ \hline
			conv5\_3 & 64 \\ \hline
			fc6 & 96  \\ \hline
			fc7 & 96  \\ \hline
			fc8 & 77  \\ \hline
        \end{tabular}
        \label{tab:vgg_pr}
  }
  \caption{Compressed AlexNet and VGG-16 models.}
\end{table}

%
%
\subsection{Blocking results}

Our first set of experiments is aimed to study the effect of variation of  block sizes on the inference time (both the decoding time and  
the computation time) for individual layers corresponding to the different configurations of the  compressed models. 
 Figure~\ref{fig:blck_f1} and ~\ref{fig:blck_f2}
show the decoding time, computation time and total time, with different block sizes for FC6 layer of AlexNet and VGGnet,
using batch size of 16. The models used for these runs  correspond to the conventional configuration. All these experiments employ MKL with 4 threads
for computation.

We observe that for very small block sizes,  the decoding and the computation time are pretty high due to overhead of the too many function calls.
For very large block sizes, the level of parallelism gets limited, leading to increase in the inference time.
Figure~\ref{fig:blck_f3} and ~\ref{fig:blck_f4} show the same charts with batch size of 256. We note that for smaller batch size, the total time is dominated by the decoding time, 
whereas the computation time takes over at larger batch sizes. However the nature of variation of  inference time with the block size is consistent across batch sizes.
We observe similar nature of plots for other configurations and batch sizes as well.

\begin{figure*}[!tbp]
  \centering
  \subfloat[AlexNet Batch size16]{\includegraphics[width=1.7in]{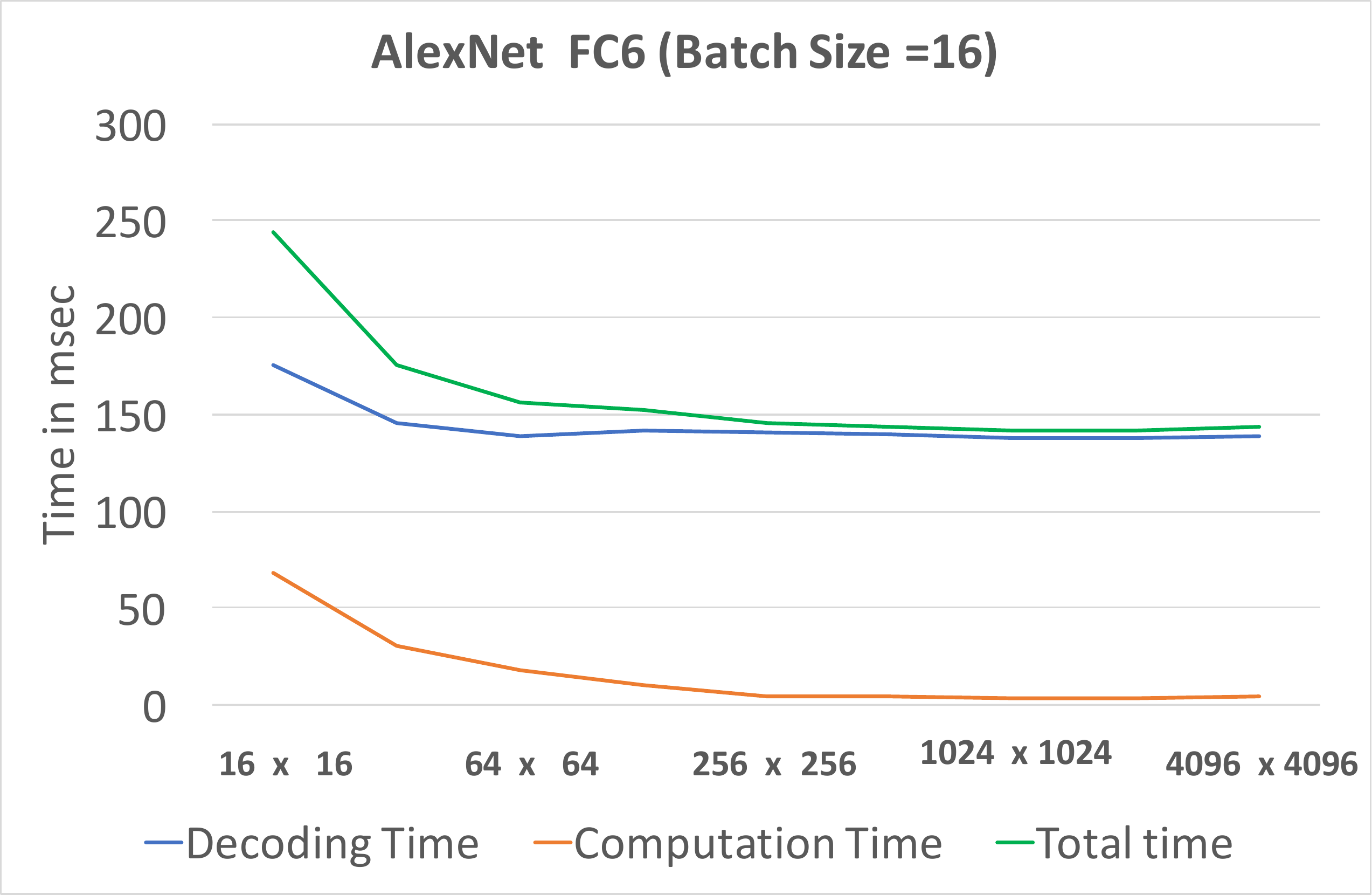}\label{fig:blck_f1}}
  \hspace{1mm}
  \subfloat[VGG-16 Batch size 16]{\includegraphics[width=1.7in]{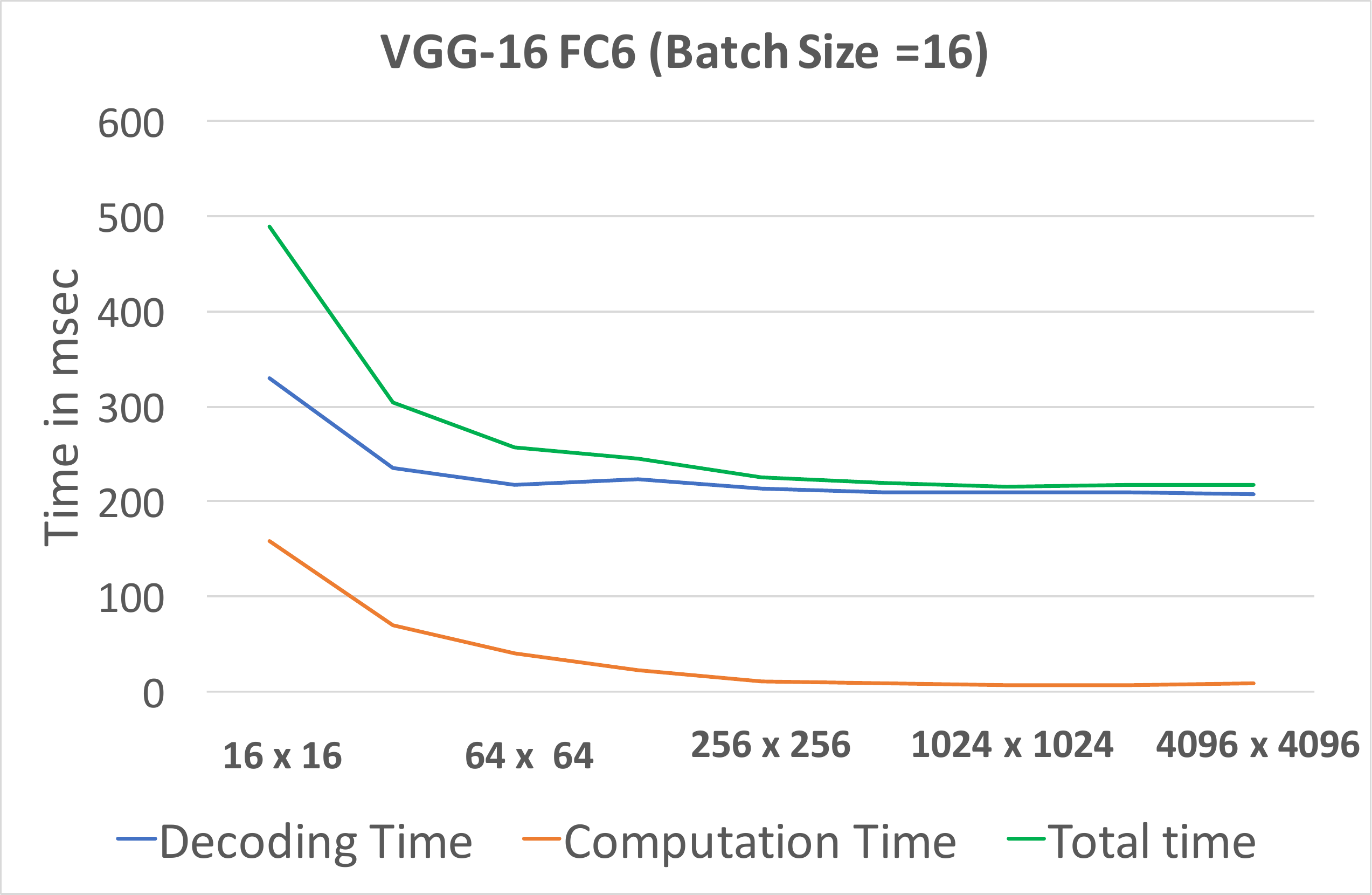}\label{fig:blck_f2}}
  \hspace{1mm}
  \subfloat[AlexNet Batch size 256]{\includegraphics[width=1.7in]{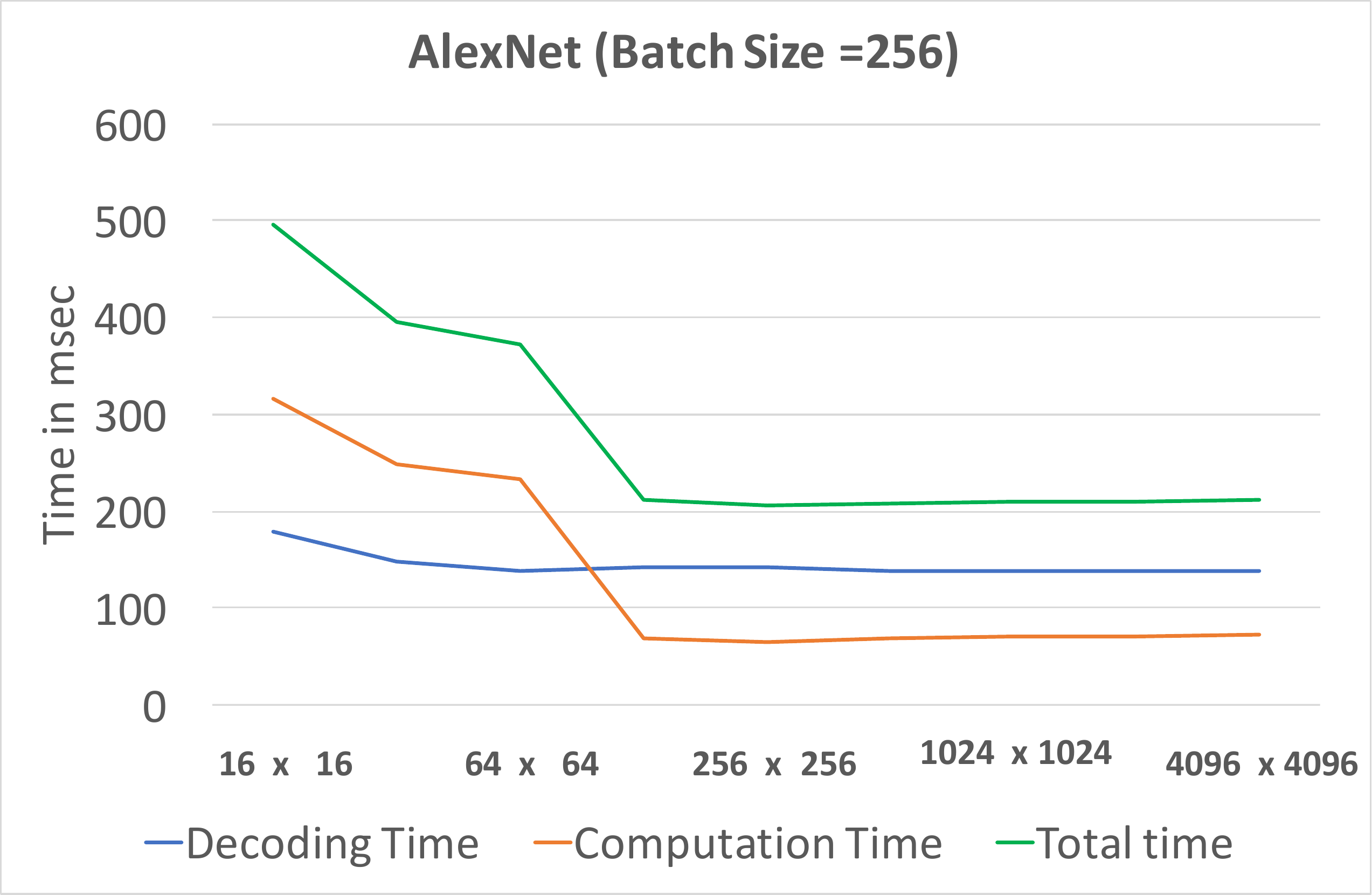}\label{fig:blck_f3}}
  \hspace{1mm}
  \subfloat[VGG-16 Batch size 256]{\includegraphics[width=1.7in]{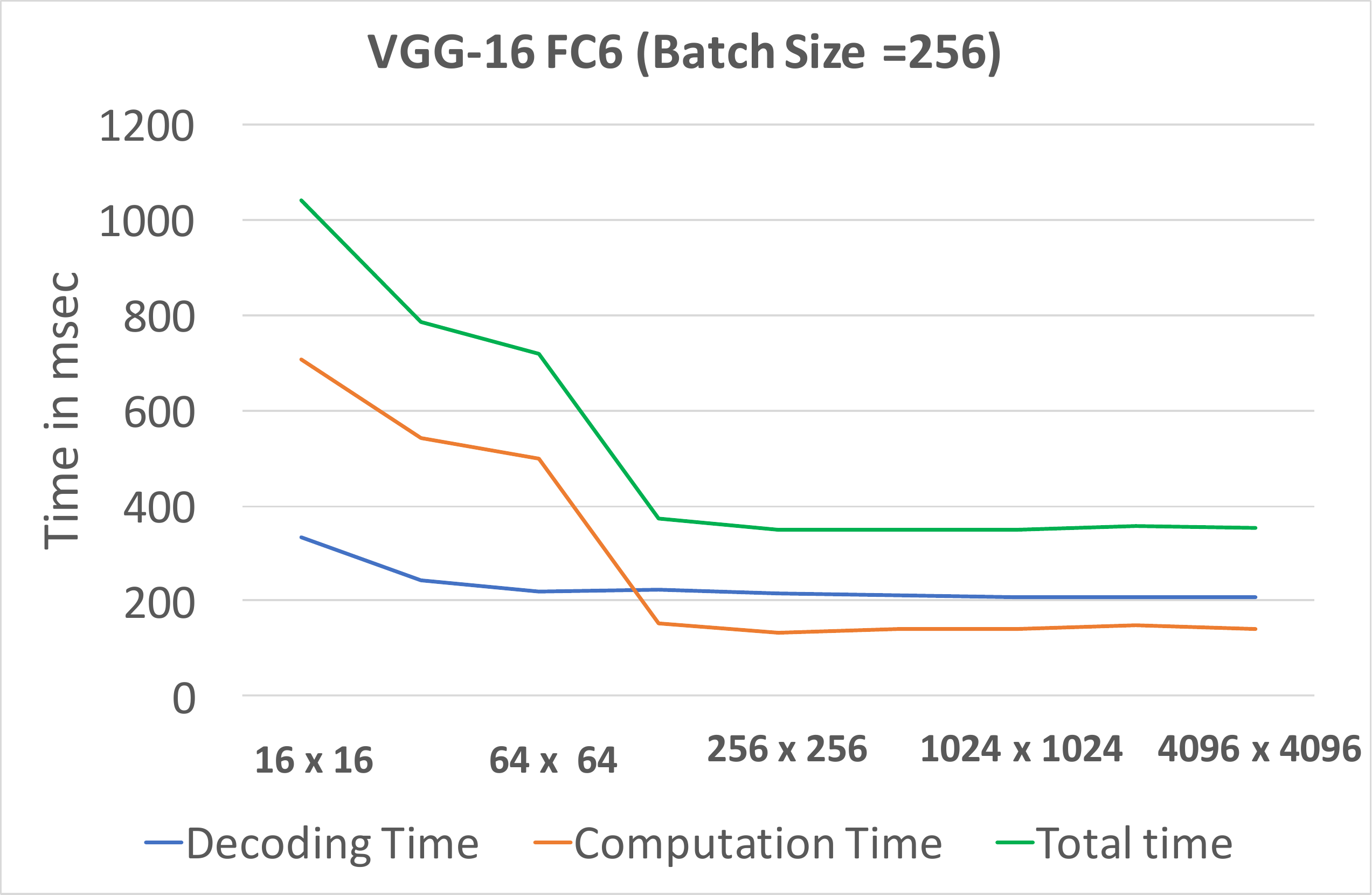}\label{fig:blck_f4}}
  \caption{Inference Time Variation with Block Size.}
\end{figure*}

We also note that the working memory increases with increase in block size. 
Table~\ref{tab:workmem} presents the working memory required for matrix matrix multiplication for FC6 layer of AlexNet and VGG-16.
Since there is not significant difference in the inference timings between block sizes in range 128 x 128 to 1024 x 1024, we fix 128 x 128 as our block size
for the subsequent experiments.

\begin{table}[h!]
\centering
\begin{tabular}{|c|c|c|}
           \hline
Blocksize & AlexNet  & VGG-16 \\ \hline
  16  x   16 &    1.26KB &  0.92KB \\ \hline
  32  x   32 &    4.57KB &   3.42KB \\ \hline
  64  x   64 &   17.33KB &  12.97KB \\ \hline
 128  x  128 &   67.40KB &  50.22KB \\ \hline
 256  x  256 &  265.78KB & 197.26KB \\ \hline
 512  x  512 &    1.03MB & 781.52KB \\ \hline
1024  x 1024 &    4.11MB &   2.98MB \\ \hline
2048  x 2048 &   14.76MB &  11.42MB \\ \hline
4096  x 4096 &   36.88MB &  42.38MB \\ \hline

\end{tabular}
\caption{Working Memory Requirement for FC6 layer}
\label{tab:workmem}

\end{table}

We next observe  the variation of activation memory requirement and the inference time with batch sizes. Table~\ref{tab:batch}   presents
the results for batch sizes of 16 and 256. Clearly, for a fixed batch size,  the activation memory required by the convolution layers is  more than  that of the 
fully-connected layers.  Inferencing applications on a low resource system generally come with a cap on the available memory. 
Suppose we consider a fictitious scenario where the maximum available memory is 20MB. 
From Table~\ref{tab:batch}, it makes sense to run the fully connected layers with batch size 256, since the memory required is well below the
permissible threshold, and there is significant increase in throughput if we process in batch of 256. 
For the convolution layers, however, processing in batch of 256 is not a desirable option because of the large memory overhead.
This motivates us to use different batch sizes for different layers during the inferencing. We present this in more detail in the next section.

\begin{table}[h!]
\centering
\begin{tabular}{|c|c|c|c|c|}
\hline
      & \multicolumn{2}{c|}{Memory (MB)} & \multicolumn{2}{c|}{Time (ms)} \\ \hline
Layer  & batch-size & batchsize & batchsize& batchsize \\
 &16 & 256 & 16 & 256 \\ \hline
conv1  &  17.72 & 283.59 &  349.93 &  5408.93 \\ \hline
norm1  &  17.72 & 283.59 &  98.87 & 1597.83 \\ \hline
pool1  &  4.27 &  68.34 & 11.68 & 176.42 \\ \hline
conv2  &  11.39 & 182.25 &  341.72 &  5745.49 \\ \hline
norm2  &  11.39 & 182.25 &  68.06 & 1081.80 \\ \hline
pool2  &  2.64 &  42.25 & 7.12 &  116.49 \\ \hline
conv3  &  3.96 &  63.38 & 153.11 &  2573.47 \\ \hline
conv4  &  3.96 &  63.38 & 204.01 &  3135.62 \\ \hline
conv5  &  2.64 &  42.25 & 135.66 &  2242.94 \\ \hline
pool5  &  0.56 &  9.00 &  1.92 &  25.72 \\ \hline
fc6  &  0.25 &  4.00 &  51.77 & 112.62 \\ \hline
fc7  &  0.25 &  4.00 &  21.06 & 46.61 \\ \hline
fc8  &  0.06 &  0.98 &  9.66 &  22.61 \\ \hline

\end{tabular}
\caption{Memory Requirement and Inference time for AlexNet individual layers}
\label{tab:batch}
\end{table}

\subsection{Inferencing with Variable Batch Size}
\label{sec:dp}
\newcommand{\OUT} {{\rm OUT}}
\newcommand{\IN} {{\rm IN}}
\newcommand{\WS} {{\rm WS}}
\newcommand{\Time} {{\rm Time}}
\newcommand{\OPT} {{\rm OPT}}
\newcommand{\TOT} {{\rm TOT}}

It is clear from the results shown in the previous section that  using a larger batch for inferencing increases the 
throughput as computing
resources are utilized more efficiently.
However, an issue with inferencing larger batches is the increase
in inferencing latency 
(due to wait time while assembling a batch, and because
larger batches take longer to process). 
Moreover the memory requirement for the input and the output
activations and buffer memory also increases for larger batch size.
Thus applications work with large batch sizes while keeping
the latency and memory utilization within certain thresholds.
The problem becomes more challenging since the 
available memory varies dynamically depending on the system load;
hence the  batch size for achieving the maximum throughput can be figured out only at the time of inferencing.
Moreover, the memory requirement and the computation time for 
inferencing varies with the layers even for a fixed batch size. 
Thus it might be advantageous to do the inferencing using different 
batch sizes for different layers.
We address this issue by proposing a dynamic programming based algorithm
for determining variable batch sizes for different layers for efficient
inferencing.
We describe our dynamic program below.

\subsection{Dynamic Programming}
Let $L_1$, $L_2$, $\cdots$, $L_f$ denote the layers of the DNN. 
For $i$ = 1,  2, $\cdots$, $f$, let \Time$(i, B)$ denote the time required to perform the inferencing computations for layer $L_i$
of the DNN using a batch size of $B$.  
Next, we let \IN$(i,B)$ and \OUT$(i,B)$ respectively denote 
the input activation and output activation memory required
to perform inferencing of layer $L_i$ with a batch size of $B$. 
Further, let \WS$(i)$ denote the size of the temporary workspace  required for layer $L_i$ computations (for instance this includes
the buffer memory required to decode blocks of the connection matrix for $L_i$). 
All the values \IN$(i,B)$,  \OUT$(i,B)$, \WS$(i)$ and \Time$(i, B)$
are obtained once for a given compressed model.
Note that the total memory required to perform inferencing computations for layer $L_i$ with a batch size of $B$ is captured by
$$ \IN(i,B) + \WS(i) +  \OUT(i,B). $$
Let {\TOT} denote the total memory available for performing the inference
computations for the entire model.

We now describe the dynamic program to determine the optimal batch size
to be used at all the individual layers in order to maximize the overall
throughput of the inferencing.
For this we define a configuration:
a configuration is a tuple $\langle i, B, A \rangle$,
where $i$ denotes the layer $L_i$, $B$ denotes a batch size and 
$A$ denotes amount of memory.
We maintain a dynamic program table {\OPT}. 
An entry \OPT$(i, B, A)$ of the dynamic program denotes the minimum time to perform the inferencing computations for layers $L_1$-$L_i$, when a batch size of 
$B$ is used for layer $L_i$, and $A$ units of memory (out of {\TOT}) are not available for performing the inferencing computations for layers $L_1$-$L_i$ (this memory is reserved for performing inferencing computations from layer $L_{i+1}$ to layer $L_f$).
Thus, we only have available ({\TOT} - $A$) units of memory
for inference computations of layers $L_1$ to $L_i$.

We say that configuration $\langle i,B,A \rangle$ is {\em feasible} if
the total memory required for performing inferencing computations at
layer $L_i$ with a batch size of $B$ is within the available memory bound,
i.e.,
$$A + \IN(i,B) + \WS(i) +  \OUT(i,B) \leq \TOT.$$

We now describe the recurrence relation for computing the 
entries of the dynamic programming table \OPT$(\cdot, \cdot, \cdot)$.
For simplicity, we assume that for every $i$, the batch size used for inferencing computations at layer $L_{i-1}$ is no more than
the batch size used for the inferencing computations at layer $L_i$.
Clearly,  \OPT$(i, B, A)$ can be finite only if  $\langle i,B,A \rangle$ is feasible.
Suppose that layer $L_i$ is computed with batch size $B$ and layer $L_{i-1}$ with a batch size $b$. 
For simplicity, we consider all $b \leq B$ such that $b$ divides $B$.
For a given $b$, 
the inferencing computations for layer $L_{i-1}$
will be performed in $ (B/b)$ phases, wherein in each phase
a batch of size $b$ will be processed up to layer $L_{i-1}$.
After the end of these phases, the $B$ output activations of 
layer $L_{i-1}$ will be fed as input activations to layer $L_i$.
Note that before the processing of the last of these phases,
\IN$(i,B-b)$ amount of output activation need to be buffered. 
Thus the total memory available for processing up to
layer $L_{i-1}$ gets reduced by \IN$(i, B-b)$ as this is required
for storing the activations before processing layer $L_i$.

We are now ready to present the recurrence relation.
For any $i > 1$, 
\begin{equation*}
\begin{split}
\OPT(i, &B, A)   =  \Time(i,B) \quad \quad  +   \\
 &\quad  \min \limits_{b \leq B} \left \{  (B/b) * \OPT(i-1,b,A+\IN(i,B-b)) \right \}\\
 &\quad \mbox{subject to} \quad \quad  \text{ $\langle i,B,A \rangle$ is feasible}
\end{split}
\end{equation*}

For the base case i.e., for $i$ = 1.
\begin{equation*}
  \OPT(1, B, A) =\begin{cases}
     \Time(1, B), & \text{if $\langle 1,B,A \rangle$ is feasible}.\\
    \infty, & \text{otherwise}.
  \end{cases}
\end{equation*}

The maximum throughput for the inferencing is obtained 
by considering the configuration that yields the 
minimum inference time per input which is 
$$\min \limits_{ B} \frac{ \OPT(f, B, 0)}{B}. $$

The above dynamic program can be easily extended to ensure that the latency of inferencing is always less than some specified threshold. In the recurrence relation,
if \OPT$(i, B, A)$ exceeds the threshold value for some $i$, $B$, and $A$, we make   \OPT$(i, B, A)$ $\leftarrow$ $\infty$. This makes sure that our optimal solution 
never has larger latency.

\subsection{Additional Storage and Computation Overhead}
The table \OPT$(\cdot, \cdot, \cdot)$ needs to be evaluated for each entry in order to figure out the individual layer batch sizes that maximise the overall throughput.
We  now figure out the additional storage and computation that is needed for the dynamic programming space complexity for standard networks like AlexNet. 
The total number of layers in AlexNet  is 14. For requested input count of 64, we consider batch sizes 
in range 1 to 64 for the second dimension. For the case where the additional available memory is twice of the model size, the third dimension is considered from 0 to 14MB
in steps of {100KB}. Thus the total size of the table is around 500KB.
Each entry computation of the table computes the minimum over a set of possible batch sizes. Thus the computation complexity is at most $\cal B$ times the size of the table,
where $\cal B$ is the maximum number of distinct batch sizes considered.

We begin our  inferencing  with a pre-processing step, which computes the  individual layer batch sizes that maximise the overall throughput using the
above dynamic program.  The actual inferencing uses the batch sizes outputted from the dynamic program.

\section{Experimental Results with Batch Size}
\label{sec:expt2}
In this section, we validate the results of our dynamic programming algorithm on practical test cases with AlexNet model.
Suppose the user requests for inference of a set of $K$ images, and is interested to get the maximum throughput for the inference.
We consider the scenarios where the total  memory in the system (in addition to the model) is 1.5x, 2x and 2.5x  times the model size.
Our baseline is selecting a fixed batch size such that (i) running any layer of inferencing using that batch size does not violate the
memory constraints (ii) out of all possible batch sizes which satisfy	 (i),  the baseline returns the batch size with maximum throughput.
We compare this baseline from our dynamic programming output, which uses variable batch sizes for different layers.
We perform our experiments $K =  32, 64, 128$ and with all the four configurations of AlexNet model (conventional pruning and
70\%, 80\%, 90\% pruning).
Figure~\ref{fig:real_dp1} - ~\ref{fig:real_dp3} compares the results of our dynamic program algorithm with the baseline 
(fixed batch size) output for AlexNet with conventional pruning. The x-axis shows the additional memory available (w.r.t to the model size)
over the model, and the y-axis plots the total time to infer K images. Our results show that the dynamic programming approach
improves the throughput  by 15-25\% over the fixed batch size approach.

\begin{figure*}[!tbp]
  \centering
  \subfloat[K=32]{\includegraphics[height = 1.3in, width=2in]{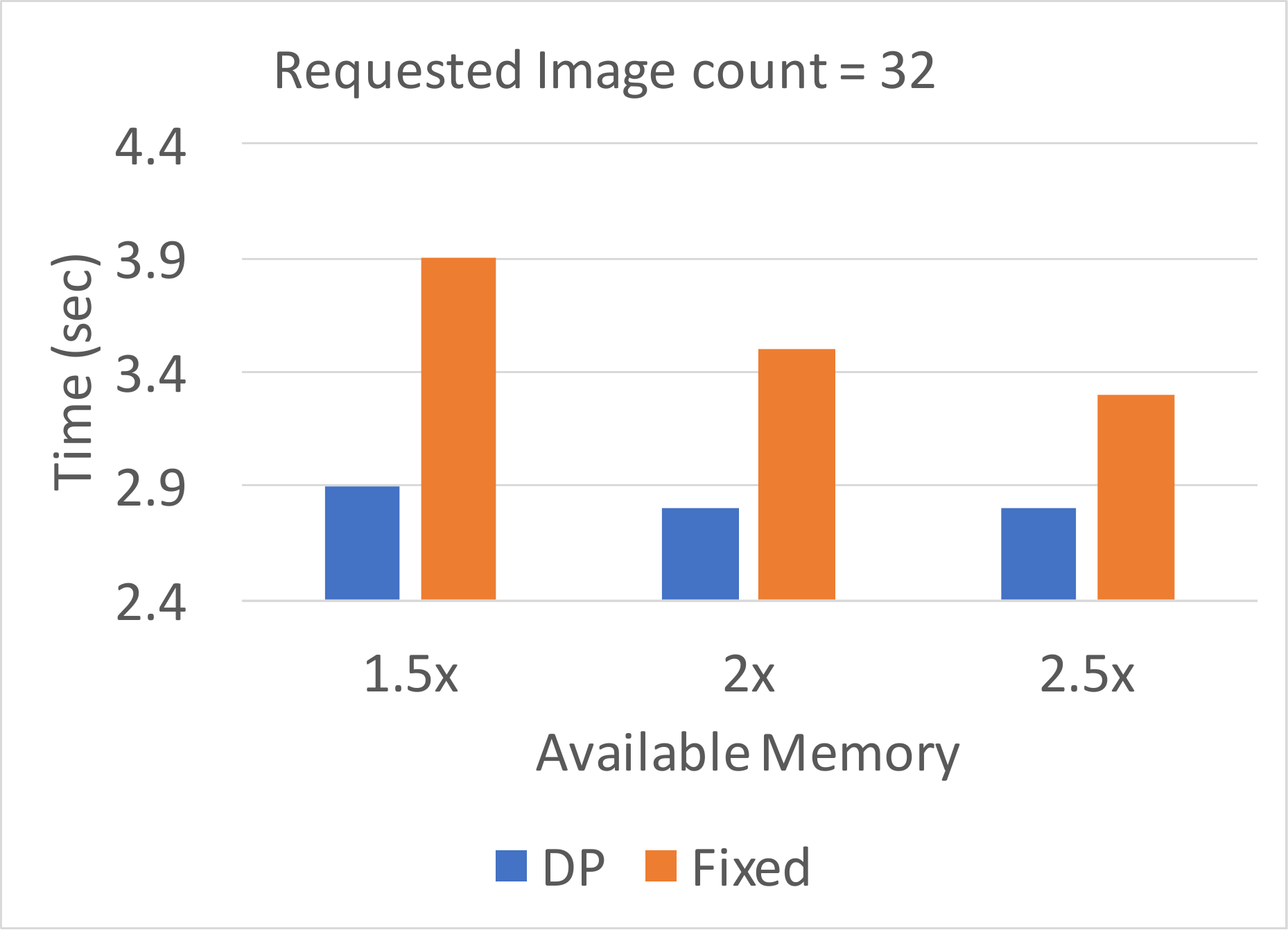}\label{fig:real_dp1}}
  \hspace{3mm}
  \subfloat[K=64]{\includegraphics[height = 1.3in, width=2in]{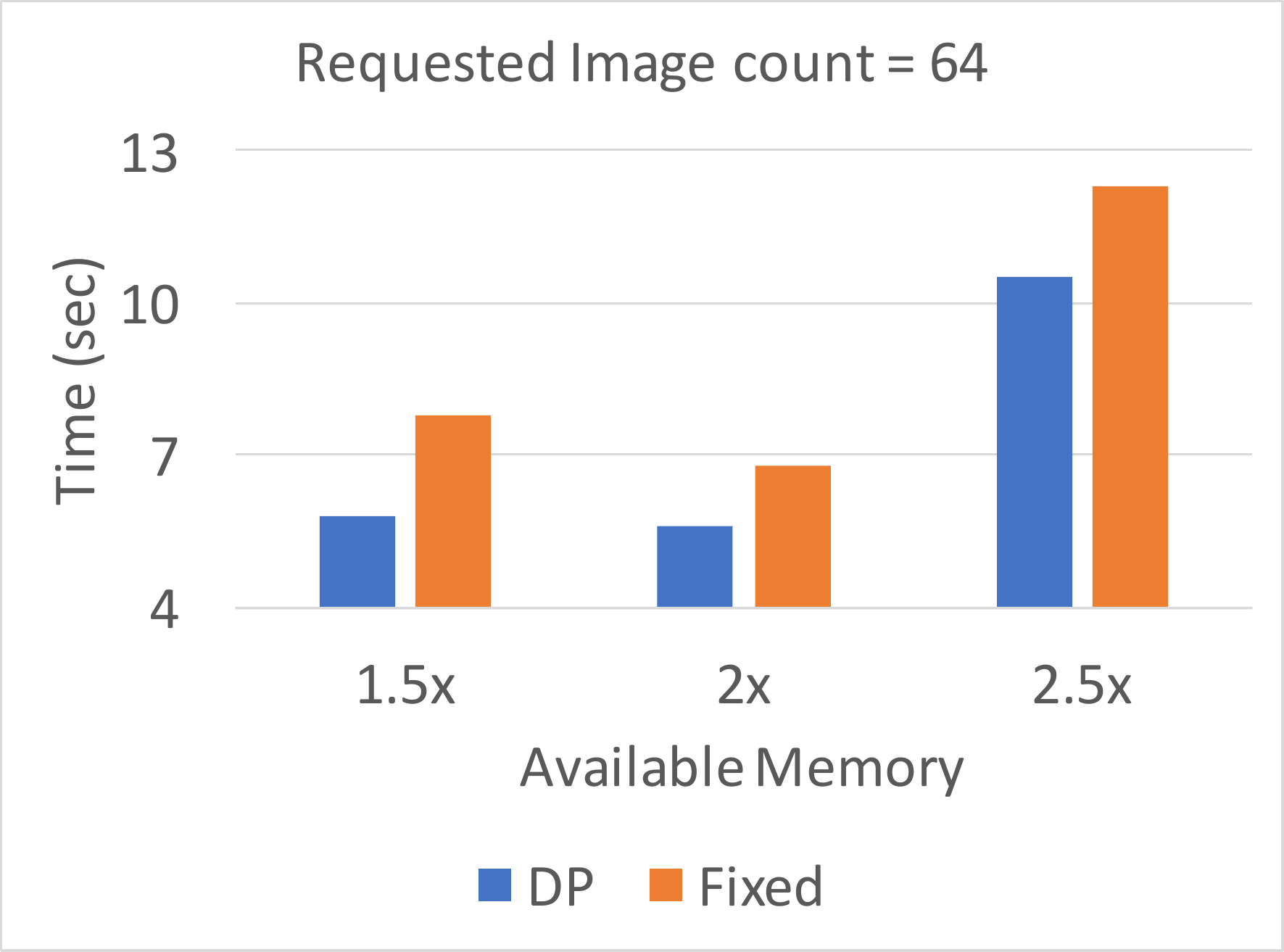}\label{fig:real_dp2}}
  \hspace{3mm}
  \subfloat[K=128]{\includegraphics[height = 1.3in, width=2in]{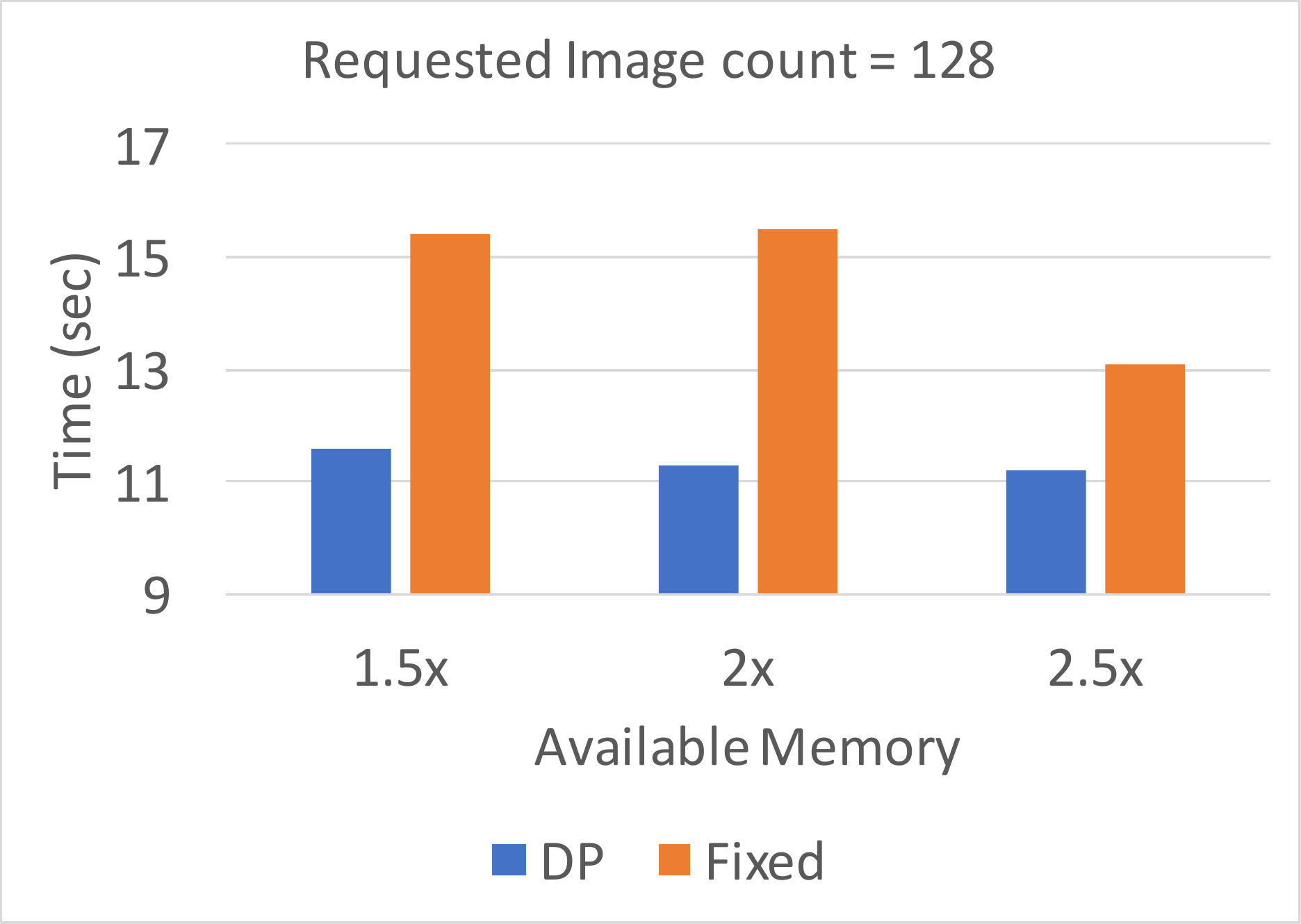}\label{fig:real_dp3}}
  \caption{Fixed batch size (baseline) time vs Time outputted from Dynamic Programming for AlexNet with conventional pruning.}
\end{figure*}

\begin{table}[h!]
\centering
\begin{tabular}{|c|c|c|c|}
\hline
 Layer & 1.5x & 2x & 2.5x \\ \hline
 conv1 & 2 & 4 & 6    \\ \hline
 norm1 & 4 & 4 & 6    \\ \hline
 pool1 & 4 & 4 & 6    \\ \hline
 conv2 & 4 & 4 & 6    \\ \hline
 norm2 & 4 & 4 & 6    \\ \hline
 pool2 & 4 & 4& 6    \\ \hline
 conv3 & 4 & 4 & 6    \\ \hline
 conv4 & 4 & 4 & 6    \\ \hline
 conv5 & 4 & 4 & 6    \\ \hline
 pool5 & 4 & 4 & 32    \\ \hline
   fc6 & 64 & 64 & 60    \\ \hline
   fc7 & 64& 64 & 60    \\ \hline
   fc8 & 64& 64 & 60    \\ \hline
\end{tabular}
\caption{Variable batching for AlexNet.}
\label{tab:dp_path}
\end{table}

\begin{figure*}[!ht]
  \centering
  \subfloat[Pruning = 70\%]{\includegraphics[height = 1.3in, width=2in]{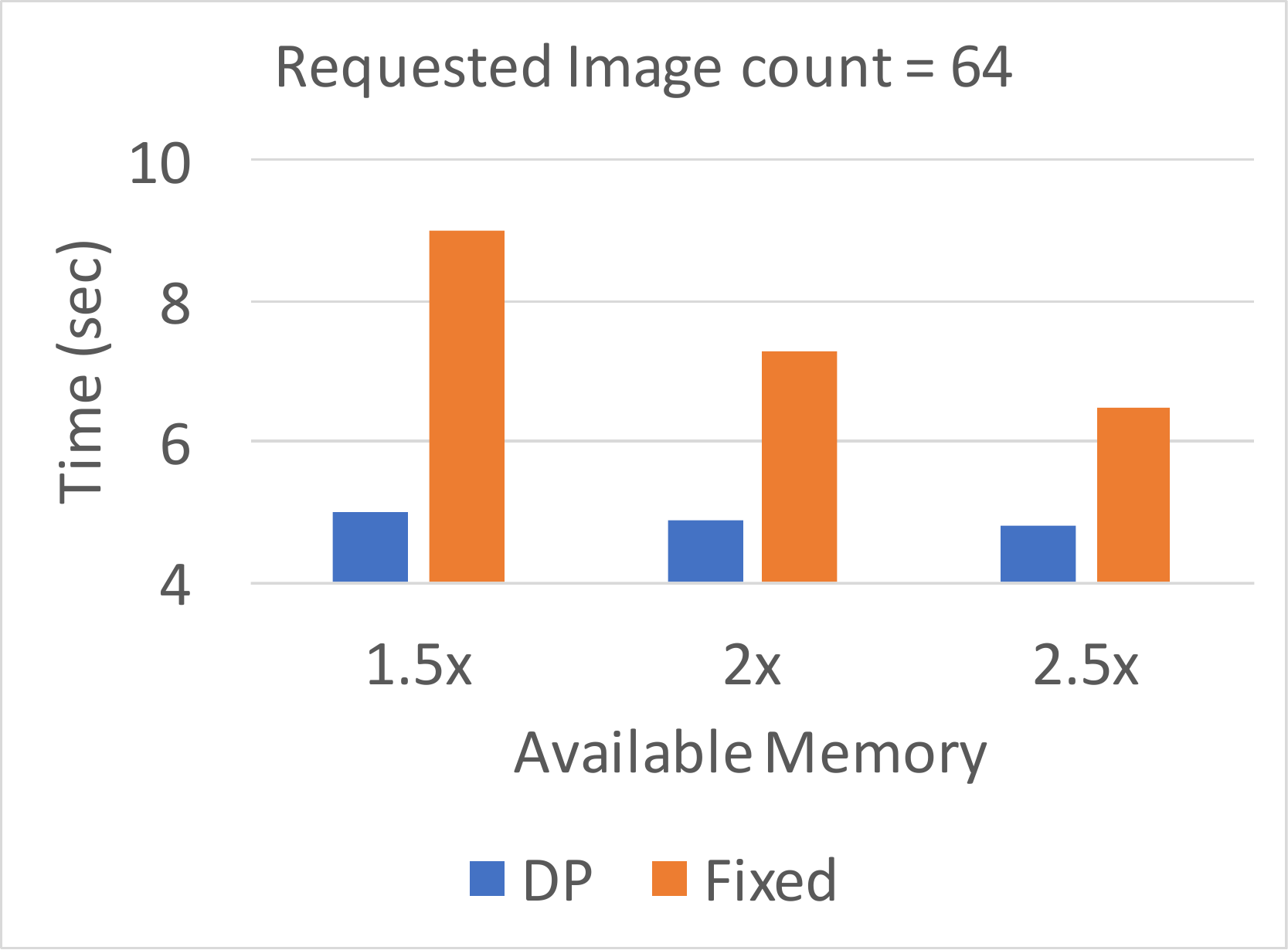}\label{fig:70_dp1}}
  \hspace{3mm}
  \subfloat[Pruning = 80\%]{\includegraphics[height = 1.3in, width=2in]{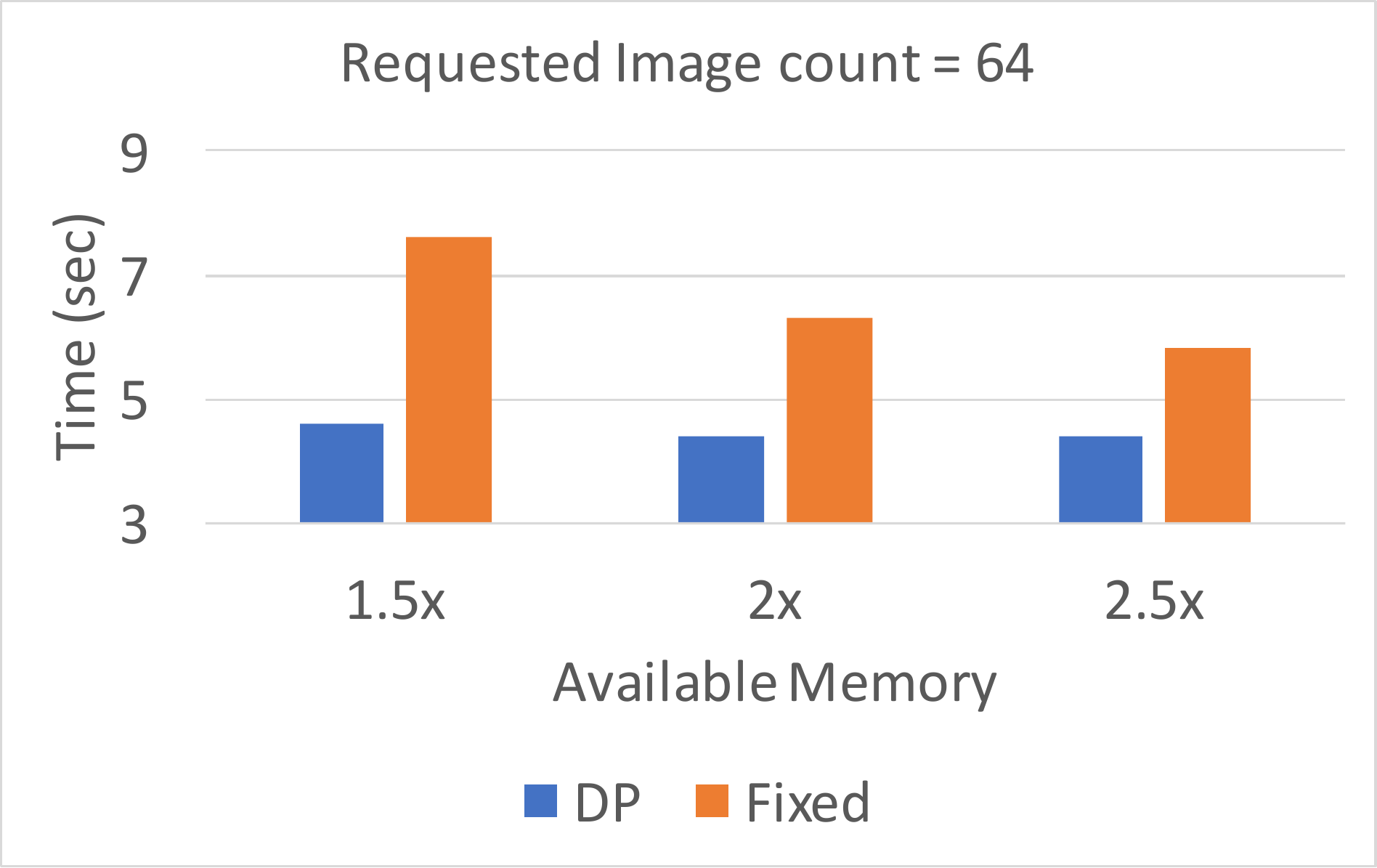}\label{fig:80_dp2}}
  \hspace{3mm}
  \subfloat[Pruning = 90\%]{\includegraphics[height = 1.3in, width=2in]{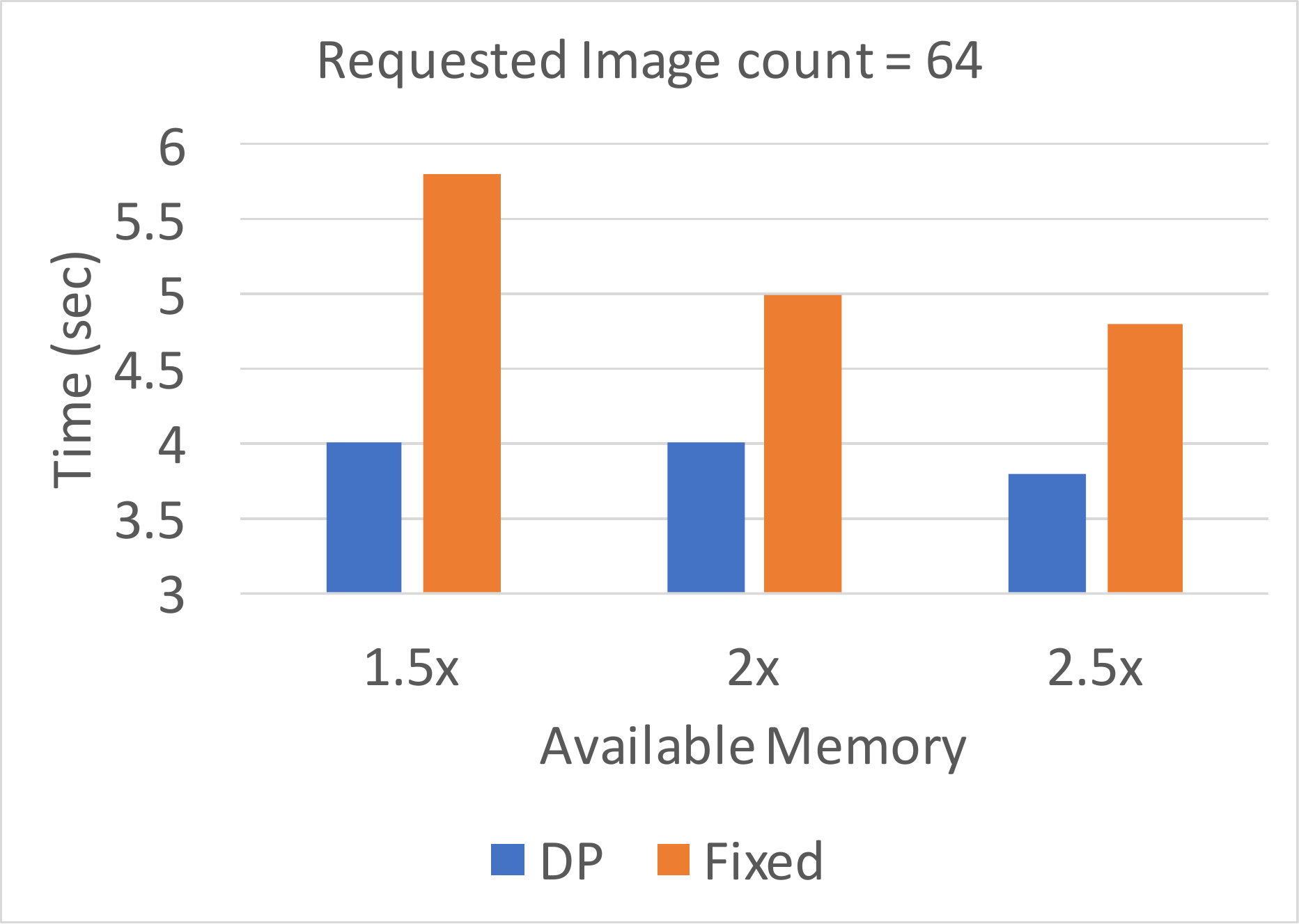}\label{fig:90_dp3}}
  \caption{Fixed batch size (baseline) time vs Time outputted from Dynamic Programming for AlexNet with 70\%, 80\% and 90\% pruning.}
\end{figure*}

Table~\ref{tab:dp_path} shows the dynamic programming output corresponding to the above run for K = 64.
 It is observed that the optimal inferencing scheme uses smaller batch sizes for the convolution layers (because of the larger memory overhead),
 and combines intermediate outputs to perform fully connected layer operations with larger batch sizes. This matches our intuition which motivated us to 
 develop the dynamic programming based solution. The dynamic programming solution  corresponding to column 2.5x picks 60 as the batchsize for final layers: thus for this case, we again compute the solution for requested input of 4, and report the total time for inferencing.  The baseline corresponding to these runs use fixed batch size of 3, 5 and 7 for  additional memory
 of  1.5x, 2x and 2.5x respectively.
 
 Figure~\ref{fig:70_dp1} - ~\ref{fig:90_dp3} extends our   results to the other configurations of the AlexNet model, namely, the
70\%, 80\% and 90\% pruned models. We show these results for fixed K of 64.  Our results show that the dynamic programming approach
performs well  over the fixed batch size approach for these scenarios as well.

\section{Concluding Remarks and Future work}
\label{sec:conc}
In this paper, we study efficient inferencing using compressed models under memory constraints. 
We propose parallel algorithms  that can use tuned math libraries available on the platform to perform inferencing efficiently
with compressed models that rely on pruning, quantization, relative indexing and encoding techniques for compression.
We study different blocking  schemes and study the effect of block sizes on  the layer of the network, its sparsity and the batch size used for AlexNet and VGG-16.
We observe that in a typical neural network inference, different layers have different sized activation memory required; thus it is beneficial
to use variable sized batching across different layers. We propose a novel dynamic programming based algorithm that figures 
out the optimal batch size for throughput maximization for the case where the batch size used for inferencing in individual layers
is a monotonically increasing sequence, 
i.e., where larger batch sizes  can be used for layers closer to the output. We show that our dynamic programming solution 
achieves 15-25\%   performance improvement  in the inference throughput over the solution employing fixed batch size  across layers.
A future work in this direction is to relax the assumption of monotonically increasing batch sizes. Our results are applicable in training of neural network models as well.
There has been recent effort for employing compressed  models in reducing training time: our 
techniques, e.g dynamic batching,  will be useful here for designing a faster forward phase.

\bibliographystyle{plain}
\bibliography{ms}

\end{document}